\documentclass{article}

\usepackage{PRIMEarxiv}

\usepackage[utf8]{inputenc} 
\usepackage[T1]{fontenc}    
\usepackage{hyperref}       
\usepackage{url}            
\usepackage{booktabs}       
\usepackage{amsfonts}       
\usepackage{nicefrac}       
\usepackage{microtype}      
\usepackage{lipsum}
\usepackage{fancyhdr}       
\usepackage{graphicx}       
\graphicspath{{media/}}     
\usepackage{amsmath}
\usepackage{amssymb}
\usepackage{algorithm}
\usepackage{algpseudocode}

\floatname{algorithm}{Pseudocode}
\usepackage{caption}
\usepackage{subcaption}
\usepackage{cite}
\usepackage{setspace} 
\usepackage{xcolor}
\usepackage{lineno}
\usepackage[normalem]{ulem}
\allowdisplaybreaks

\title{ARRTOC: Adversarially Robust Real-Time Optimization and Control
}

\author{
  Akhil Ahmed, Ehecatl Antonio del Rio-Chanona, Mehmet Mercangöz\thanks{Corresponding Author} \\
  Centre for Process Systems Engineering, Department of Chemical Engineering \\
  Imperial College London \\
  London\\
  \texttt{\{a.ahmed21, a.del-rio-chanona, m.mercangoz\}@imperial.ac.uk} \\
}

\begin{document}
\maketitle

\begin{abstract}
Real-Time Optimization (RTO) plays a crucial role in the process operation hierarchy by determining optimal set-points for the lower-level controllers. However, at the control layer, these set-points may be difficult to track due to challenges in implementation as a result of disturbances, measurement noise, and actuator performance limitations. To address this, in this paper, we present the Adversarially Robust Real-Time Optimization and Control (ARRTOC) algorithm. ARRTOC addresses this issue by finding set-points which are both optimal and inherently robust to implementation errors at the control layers. ARRTOC draws inspiration from adversarial machine learning, offering a novel constrained Adversarially Robust Optimization (ARO) solution applied to the RTO layer. By integrating controller design with RTO, ARRTOC enhances overall system performance and robustness by ensuring the chosen set-points are tailored to the underlying controller designs. To validate our claims, we present three case studies: an illustrative example, a bioreactor case study, and a multi-loop evaporator process. The proposed approach is found to improve RTO objectives, such as profit, by as much as $50\%$ in some case studies compared to RTO formulations which ignore the performance of the control layers. 
\end{abstract}


\section{Introduction}\label{sec:intro}

Real-Time Optimization (RTO) is a critical task for the efficient operation of a process system. It allows optimal set-points to be found for the lower-level controllers in the process operation hierarchy  while ensuring operational constraints are satisfied. All of this must be handled in real-time due to changes in costs and product specifications amongst other factors \cite{Siirola1996, Smith2016, Aydin2016, Navia2019, Patron2022, Daoutidis2024}. Once the set-points are calculated, they are sent to the lower-level control layers, where the controllers pursue three primary objectives: set-point tracking, disturbance rejection, and noise attenuation \cite{Seborg2011}. It is crucial to acknowledge that these three objectives inherently conflict with one another -- a fundamental observation from linear control theory \cite{Ogata2009}. Notably, for nonlinear systems, these conflicts are exacerbated, intensifying the challenges associated with achieving satisfactory control performance \cite{Khalil2014}. Despite this, all objectives must be adequately satisfied while ensuring the controller adheres to tight time constraints. Consequently, when designing controllers, a trade-off is usually sought. For example, for a Model Predictive Controller (MPC), while it may be possible to design a fully nonlinear MPC solution that is robust to certain disturbance and noise characteristics and other forms of uncertainty, this may not be practically implementable \cite{Paulson2020, Qin2003, Henson1998, Charitopoulos2021, Mayne2000}. 

Importantly, traditionally selected set-points can be sub-optimal if control layer implementation errors (such as disturbances and noise) are not explicitly accounted for, especially if the process is disturbance-sensitive and highly nonlinear \cite{Jaschke2017}. In this case, the control layer assumes a heightened responsibility for actively rejecting disturbances and attenuating noise, thereby inevitably placing greater demands on the resource-constrained controllers \cite{Jaschke2017}. This is the challenge that we address in this paper. Our approach leverages the interaction between the upper-level RTO layer and the lower-level control layers. Instead of confining RTO's role solely to identifying optimal controller set-points, we advocate for an approach in which the RTO layer selects set-points that are not only optimal but also inherently robust to control layer implementation errors. The chosen set-point provides a form of passive disturbance rejection and noise suppression, thereby reducing the burden on the control layers. This is achieved by incorporating elements from adversarial machine learning \cite{Bai2021, Silva2020, Carlini2019}. In adversarial machine learning, perturbations are crafted to deceive models into producing incorrect outputs. Similarly, in the context of RTO and control, disturbances and noise entering the control layer can be viewed as adversarial perturbations that hinder control system performance. The presence of these adversarial perturbations in both domains motivates the use of Adversarially Robust Optimization (ARO) techniques to mitigate their effects. In RTO, this entails identifying set-points deliberately designed to withstand various sources of implementation errors arising from the control layers, allowing the controllers to operate in a stable and consistent manner. This is achieved using our practical constrained adversarially robust optimization algorithm applied to the RTO layer, termed the Adversarially Robust Real-Time Optimization and Control (ARRTOC) algorithm. ARRTOC accomplishes this by addressing both optimality and operability concerns simultaneously. The resulting set-point is intentionally designed to be insensitive to potential disturbances and noise, effectively offloading part of the responsibility of robustness from the controller to the RTO layer. This allows us to use more computational time resources at the RTO level (where they are abundant) and avoid this at the control level (where they are lacking). 

\subsection{Related Work}\label{sec:related_works}

The identification of operable design regions to achieve robustness to implementation errors in fields such as RTO, control or process design presents a significant challenge \cite{Gazzaneo2020, Jaschke2017, Zhang2023}. Broadly speaking, most methods involve computationally-intensive sensitivity-based approaches that are typically prone to unverified assumptions (e.g. the correctness of the underlying model) and can suffer from the curse of dimensionality, rendering these methods impractical \cite{Gazzaneo2020, Georgiadis2002, Lima2010, Dinh2023, Sakizlis2004}.

Within the realm of RTO specifically, while robust RTO algorithms have been proposed, these have primarily emphasised robustness against modelling errors and uncertainty in the steady-state input-output map \cite{Marchetti2011, Janaqi2013, MacKinnon2023}. However, these approaches have overlooked the critical aspect of robustness to implementation errors at the control layer. In other words, they tend to consider RTO and control independently, disregarding the underlying controller design. This results in set-points which, when implemented at the control layers, can be difficult to operate around or are entirely inoperable. Although researchers recognise the importance of this problem, they have typically resorted to the aforementioned sensitivity-based methods \cite{Gazzaneo2020, Georgiadis2002}. As an alternative approach, the idea of "self-optimizing control structures" relegates the implementation error robustness problem to the control layers \cite{Jaschke2017}. In this paradigm, the problem is framed as selecting the "best" set of controlled variables that minimize the impact of implementation errors on the RTO objective function \cite{Jaschke2017, Skogestad2000}. Although this approach works well for most cases, it might still run into two main difficulties. First, even with the optimal choice of controlled variables, the system may be too sensitive to implementation errors due to strong non-linearities. Second, the choice of controlled variables may not be a practical or feasible choice. Instead, as noted by \cite{Skogestad2000}, the solution of this problem would benefit from a formal robust treatment of implementation errors at the RTO layer directly. Our paper, aims to address this gap in the literature through the use of ARO.

The fundamental issue with a formal robust treatment of implementation errors at the RTO layer lies in the possible complexity of the robust RTO problem. Any proposed RTO algorithm must accommodate various problem settings, including the full online setting (e.g. changing prices or plant-model mismatch), scenarios with unknown or partially known objective function and constraints, and even scenarios with implicitly given simulation-based objective and constraints. This flexibility is crucial, especially in industrial settings where process models are derived from in-house or proprietary process simulation software \cite{Liporace2009}.

Unfortunately, as explored by \cite{Bertsimas2010}, the majority of existing robust optimization algorithms fail to address these complexities. To address this gap in the robust optimization literature, their work introduced a novel approach, which subsequently found widespread acceptance within the machine learning community (under the banner of adversarially robust optimization), where similar complexities in the problem setting are faced. This paradigm shift proved instrumental in tackling the pressing need for adversarial robustness in machine learning settings \cite{Bogunovic2018, Bai2021, Silva2020, Carlini2019}. Critically, ARO makes minimal assumptions regarding problem structure and is applicable in scenarios with simulation-based objectives making it well suited for practical engineering problems \cite{Bogunovic2018, Silva2020}. Indeed, parallels can be drawn between the need for operability in RTO and control with the need for adversarial robustness in machine learning. To this end, our work builds upon the foundation laid by ARO and capitalizes on the similarities between RTO operability and adversarial robustness in machine learning to introduce the ARRTOC algorithm.

\subsection{Objectives and Contributions}\label{sec:contributions}

Given the above, the objective of our paper is to introduce our algorithm ARRTOC which applies a constrained ARO algorithm to the RTO layer to account for control layer implementation errors. This algorithm addresses an open challenge in the literature as reviewed in this section. Specifically, in the first case, we avoid the use of computationally expensive and possibly impractical sensitivity-based methods which are typically used to select operable designs/set-points. Secondly, while self-optimizing control can be used to select the best controlled variables, this may not be sufficient by itself nor practically implementable. Instead, a formal robust treatment of implementation errors at the RTO layer is required. However, existing robust algorithms are not able to accommodate the possible complexities which may be encountered at the RTO layer. On the contrary, advances in adversarial machine learning and specifically, adversarially robust optimization, provides a pragmatic path to robustify the RTO layer to implementation errors. Crucially, ARO in its current form fails to account for constraints which our algorithm, ARRTOC, addresses. In this paper, we aim to demonstrate the practical implementation of our algorithm, prioritising focus on its immediate application. To maintain focus on this core objective, we solve steady-state optimization problems in our case studies to identify robust set-points, which are then integrated into the lower-level control layers, to show the significant enhancements in operability. 
 
The novelty and contributions of our paper are as follows:

\begin{enumerate}
    \item \textbf{Development of a practical RTO algorithm based on Adversarially Robust Optimization}: We present a novel ARO algorithm, ARRTOC, which handles constraints and uncertainties relevant to RTO problems.
    \item \textbf{Emphasis on operability for stable and consistent results}: In contrast to solely seeking naive optimality in RTO, we recognise the importance of operability. Our research highlights that operability should be a key consideration in RTO, alongside optimality.
    \item \textbf{Leveraging the interaction between RTO and control through ARRTOC}: ARRTOC leverages ARO to counteract control layer implementation errors. This approach identifies optimal and operable set-points for the controllers, enhancing the overall system performance.
    \item \textbf{Unified and holistic balance of robustness}: Through our experiments, we demonstrate that the coupling between RTO and control offered by ARRTOC enables the discovery of set-points that strike a balanced level of robustness for both layers without risking excessive conservatism.
    \item \textbf{Loose coupling and compatibility with various RTO and controller architectures}: ARRTOC's loose coupling between the RTO and control layers accommodates diverse controller structures and designs. Moreover, it integrates with various model types (simulation-based, mechanistic, data-driven or hybrid) used at the RTO layer. This compatibility allows for the employment of important RTO model adaptation algorithms or novel controller designs in conjunction with ARRTOC, making it a versatile and adaptable solution.
\end{enumerate}

The rest of the paper is structured as follows. In Section \ref{sec:background_meth}, we present the Adversarially Robust Real-Time Optimization and Control (ARRTOC) algorithm. We start by introducing the concept of Adversarially Robust Optimization (ARO) in the context of Real-Time Optimization (RTO) and demonstrate how constraints can be incorporated within ARO to yield ARRTOC. Moving to Section \ref{sec:results_discussion}, we apply ARRTOC to three case studies, progressively increasing in complexity. Section \ref{rd:illustrative_example} presents an illustrative case study, offering an intuitive understanding of ARRTOC's principles. This groundwork then informs two practical applications: Section \ref{rd:bioreactor} addresses wash-out in continuous bioreactors using ARRTOC. Lastly, Section \ref{rd:evaporator} applies ARRTOC to a multi-loop evaporator case study. Concluding in Section \ref{sec:conclusion}, we summarise the paper's findings and suggest potential avenues for future research. 

\section{Background and Methodology}\label{sec:background_meth}

In this section, we delve into the ARRTOC algorithm, contextualising it within the ARO framework applied to RTO. We do this by providing a self-contained tutorial-style introduction to ARRTOC, leveraging insights from a practical ARO algorithm which forms the core of our algorithm \cite{Bogunovic2018, Bertsimas2010}. As will be noted, most ARO algorithms primarily address unconstrained problems, a design choice that resonates with its primary application in adversarial machine learning, a domain where constraints are typically absent \cite{Bertsimas2010, Bogunovic2018, Bai2021, Silva2020, Carlini2019}. Of course, for RTO applications, constraints are vital as they may be safety-critical \cite{Liporace2009}. To address this, we begin by defining the constrained problem in section \ref{constrained:prob_def}, emphasising the concept of adversarial robustness and its relevance to RTO. Next, in section \ref{constrained:intuition}, we provide an intuitive overview of the algorithm, focussing on how ARRTOC handles constraints. This overview leads to a high-level pseudocode representation, which we comprehensively explore in sections \ref{constrained:step_1}, \ref{constrained:step_2}, \ref{constrained:step_3a} and \ref{constrained:step_3b}. 

\subsection{Problem Definition}\label{constrained:prob_def}

Consider the nominal constrained real-time optimization problem as defined below:

\begin{equation}\label{eq:nom_opt_constrained}
\begin{aligned}
\underset{\mathbf{x}}{\text{min}} &\; f(\mathbf{x})\\
\text{s.t.} &\; h_j(\mathbf{x}) \leq 0
\end{aligned}
\end{equation}
where $\mathbf{x} \in \mathbb{R}^{n_x}$ is the decision variable, which in the context of RTO will be the set-point for the controllers and $f : \mathbb{R}^{n_x} \to \mathbb{R}$, the objective function, typically reflects operating goals, such as economic performance metrics while $h_j : \mathbb{R}^{n_x} \to \mathbb{R}$, is a constraint function, which may be safety critical in the RTO setting. For the case studies in section \ref{sec:results_discussion}, we assume the objective function and constraints are known apriori (i.e. the underlying system model is known), however, this may not be true in most RTO problems which require an online approach. As noted in section \ref{sec:intro}, we do this to focus on the primary goal of the paper - to demonstrate the applicability of ARO to RTO and control problems in order to account for control layer implementation errors at the RTO layer. However, as will be shown in this section, no assumptions are made regarding the functional form of $f(\mathbf{x})$ and $h_j(\mathbf{x})$. Indeed they may be partially or completely unknown, data-driven, mechanistic, hybrid or possibly even simulation-based, allowing the ARRTOC algorithm to be easily deployed in real-time online settings. Similar to \cite{Bertsimas2010}, we only assume cheap access to the function value and gradient.

When implementing the optimal set-point, $\mathbf{x}$, controllers face vulnerability to implementation errors, primarily stemming from disturbances and noise, as well as other sources outlined in the control literature \cite{Seborg2011, Ogata2009}. Quantifying these errors accurately is challenging, with often only approximate bounds available. Nonetheless, robustness to such errors is crucial for control objectives, such as set-point tracking, disturbance rejection, and noise suppression. As discussed in section \ref{sec:intro}, complex robust control algorithms, while theoretically feasible, may not be practical due to the time constraints imposed on the controllers. Therefore, at the RTO layer, we adopt an alternative approach: identifying set-points inherently robust to implementation errors, thereby lessening the demands placed on the resource-constrained controllers.

In our problem definition, we denote implementation errors as $\boldsymbol{\Delta}\mathbf{x} \in \mathbb{R}^{n_x}$. These are the fluctuations from the set-point due to perturbations such as disturbances and noise. Consequently, the actual realisation of a controller set-point may be $\mathbf{x} + \boldsymbol{\Delta}\mathbf{x}$ as opposed to $\mathbf{x}$. We assume the implementation error, $\boldsymbol{\Delta}\mathbf{x}$, resides within the uncertainty set, $\mathcal{U}$ defined as:

\begin{equation}\label{eq:unconstrained_uncertainty_set_pre}
\mathcal{U} = \left\{\boldsymbol{\Delta}\mathbf{x} \;| 
 \; d(\boldsymbol{\Delta}\mathbf{x}) \leq \Gamma \right\}
\end{equation}
where $d : \mathbb{R}^{n_x} \to \mathbb{R}$, is a distance function which defines the "distance" of the implementation error from the decision variable $\mathbf{x}$. In this section, we exclusively consider the scenario where $d(\boldsymbol{\Delta}\mathbf{x}) = ||\boldsymbol{\Delta}\mathbf{x}||_2$ to give the uncertainty set defined by:

\begin{equation}\label{eq:unconstrained_uncertainty_set}
\mathcal{U} = \left\{\boldsymbol{\Delta}\mathbf{x} \;| 
 \; ||\boldsymbol{\Delta}\mathbf{x}||_2 \leq \Gamma \right\}
\end{equation}
where $\Gamma \in \mathbb{R}$ defines the upper-bound for the distance from the decision variable $\mathbf{x}$. It represents the largest possible perturbation or implementation error we must safeguard against. In the context of control, this can represent the most significant anticipated disturbance-induced perturbation in the controlled state.  This value can be chosen based on controller design considerations, often determined through simulation studies and tuning, as demonstrated in our case studies in section \ref{sec:results_discussion}. Ultimately, the choice of $\Gamma$ depends on achieving the right balance between robustness at the RTO layer and the existing control layer robustness, an aspect we delve into extensively in section \ref{sec:results_discussion}.

However, while this uncertainty set is acceptable in the adversarial machine learning context \cite{Bogunovic2018} (and we use it in this section for ease of illustration), this is not the case for RTO and control problems. The set described by equation (\ref{eq:unconstrained_uncertainty_set}), describes an n-sphere, which implicitly assumes that all system states are subject to the the same maximum possible perturbation of $\Gamma$. Of course, in reality, for many systems, it is known that disturbances and noise may affect one state more than the others or more generally, the different states of the system will have different maximum possible perturbations, $\Gamma_i$. One possible solution to this might involve setting $\Gamma$ as the highest $\Gamma_i$ among states. However, this approach risks excessive conservatism. To accommodate potential asymmetry in implementation errors at the control layers, we introduce a generalisation of the uncertainty set. This set explicitly accounts for different maximum perturbations, $\Gamma_i$, for each state. Rather than an n-sphere, we define an n-ellipsoid:

\begin{equation}\label{eq:ellipsis_uncertainty_set}
\mathcal{U} = \left\{\boldsymbol{\Delta}\mathbf{x} \in \mathbb{R}^{n_x} \; \middle |  \; \sum_{i}^{n_x} \frac{{\boldsymbol{\Delta}\mathbf{x}_i}^2}{{\Gamma_i}^2} \leq 1 \right\}
\end{equation}
where $\boldsymbol{\Delta}\mathbf{x}$ are the implementation errors as before and $\Gamma_i$ defines the maximum possible perturbation or implementation error to protect against along the $i$th dimension. This change, while subtle, is powerful for RTO and control problems as it can be used to find adversarially robust set-points which are more targeted and tailored to the level of robustness required for each of the states independently. This will be demonstrated in the  case studies in section \ref{sec:results_discussion}.

We seek an adversarially robust set-point. This is a set-point which is robust to the worst possible implementation error of our process:

\begin{equation}\label{eq:robust_constrained_prob}
\begin{aligned}
\underset{\mathbf{x}}{\text{min}} & \underset{\boldsymbol{\Delta}\mathbf{x} \in \mathcal{U}}{\text{max}} \; f(\mathbf{x} + \boldsymbol{\Delta}\mathbf{x})\\
\text{s.t.} & \underset{\boldsymbol{\Delta}\mathbf{x} \in \mathcal{U}}{\text{max}} \; h_j(\mathbf{x} + \boldsymbol{\Delta}\mathbf{x}) \leq 0
\end{aligned}
\end{equation}
The problem formulation implies that a robust set-point should (i) minimize the worst-case cost as defined as $\underset{\boldsymbol{\Delta}\mathbf{x} \in \mathcal{U}}{\text{max}} \; f(\mathbf{x} + \boldsymbol{\Delta}\mathbf{x})$ and (ii) simultaneously ensure no constraints are violated for any $\boldsymbol{\Delta}\mathbf{x} \in \mathcal{U}$.

\begin{figure}
\begin{center}
\includegraphics[width=0.7\textwidth]{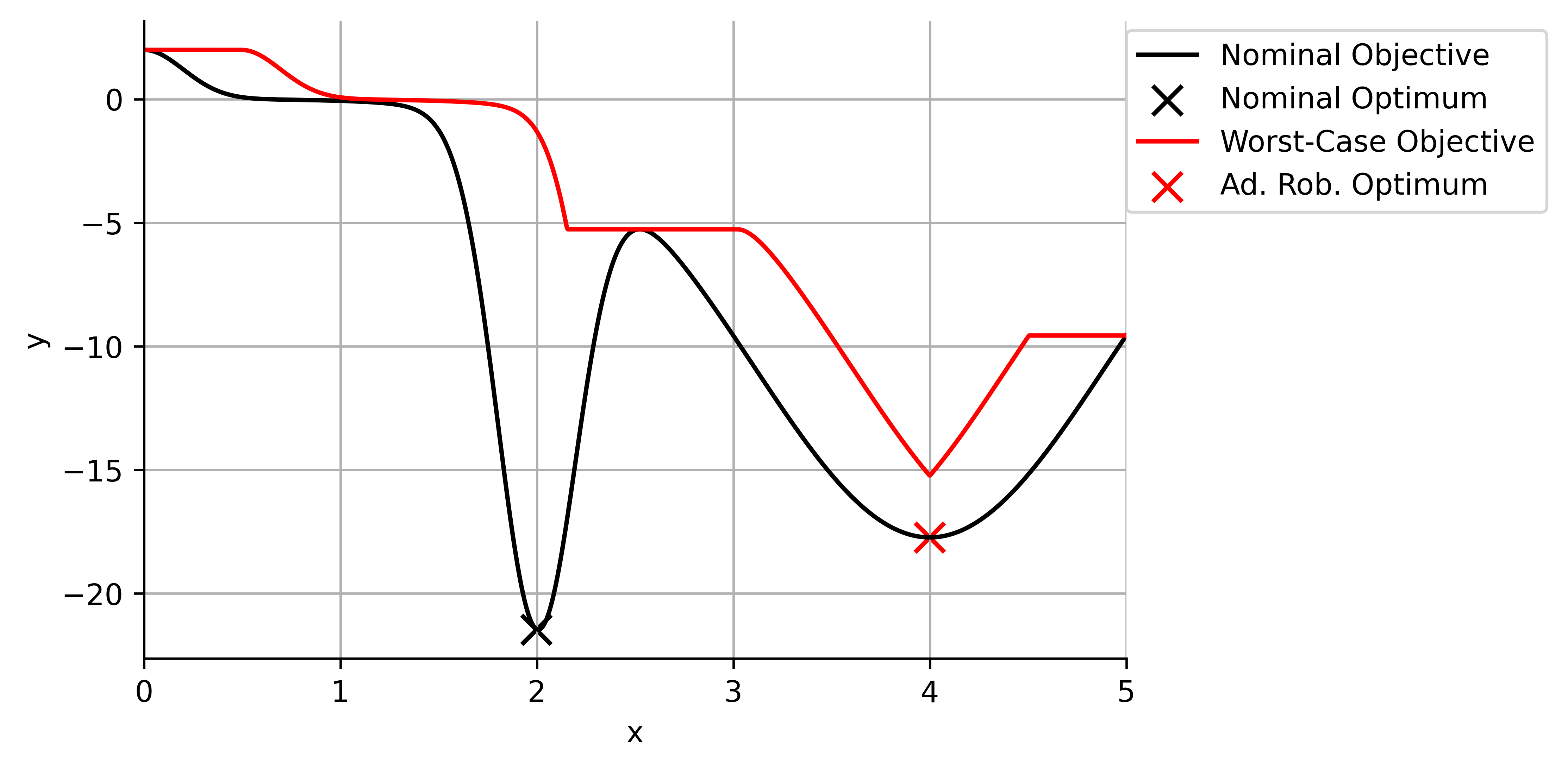} 
\caption{A simple one dimensional example depicting the nominal objective function as a black curve and the worst-case objective function ($\Gamma = 0.5$) as a red curve. The nominal optimum is depicted as a black cross and the adversarially robust optimum is depicted as a red cross.} 
\label{fig:1d_intuition}
\end{center}
\end{figure}

This concept is illustrated in Figure \ref{fig:1d_intuition} using a one-dimensional unconstrained example. The nominal objective function is in black and the worst-case objective function, $\underset{\boldsymbol{\Delta}\mathbf{x} \in \mathcal{U}}{\text{max}} \; f(\mathbf{x} + \boldsymbol{\Delta}\mathbf{x})$, is in red, with $\Gamma = 0.5$ for clarity. The black cross marks the nominal optimum, while the red cross represents the adversarially robust optimum, found by solving equation (\ref{eq:robust_constrained_prob}). The distinction is evident: the nominal optimum is susceptible to implementation errors, as shown by the significantly higher worst-case objective value compared to its nominal value at this point. Conversely, the adversarially robust optimum remains close to its nominal value even in the face of worst-case implementation errors, making it a stable choice for real-world implementation. 

While the aforementioned concepts are appealing, particularly within the domain of RTO and control, solving equation (\ref{eq:robust_constrained_prob}) 
poses a formidable challenge without simplifying assumptions that compromise practicality to allow the problem to be solved online and in real-time. Nevertheless, we outline an approach based on the work of \cite{Bertsimas2010} that yields a practical algorithm, avoiding unnecessary assumptions and offering a pragmatic solution to the robust problem in equation (\ref{eq:robust_constrained_prob}), suitable for various practical problem scenarios.

\subsection{The Intuition Behind The ARRTOC Algorithm}\label{constrained:intuition}

In this section, we offer an intuitive understanding of the ARRTOC algorithm through visual examples, which is critical to understanding our methodology. This leads to a high-level pseudocode description of the algorithm. We then enhance this initial insight and pseudocode using precise descriptions for each step, as detailed in sections \ref{constrained:step_1}, \ref{constrained:step_2}, \ref{constrained:step_3a} and \ref{constrained:step_3b}. 

Given a point, $\mathbf{\hat{x}}$, we define the neighbourhood set of this point as the set of all points, $\mathbf{x}$, which are reachable given the implementation errors defined by $\boldsymbol{\Delta}\mathbf{x} \in \mathcal{U}$:

\begin{equation}\label{eq:unconstrained_neighbourhood_set}
\mathcal{N}(\mathbf{\hat{x}}) = \left\{\mathbf{x} \;| 
 \; ||\mathbf{x} - \mathbf{\hat{x}}||_2 \leq \Gamma \right\}
\end{equation}

\begin{figure}
     \centering
     \begin{subfigure}[b]{0.3\textwidth}
         \centering
         \includegraphics[width=\textwidth]{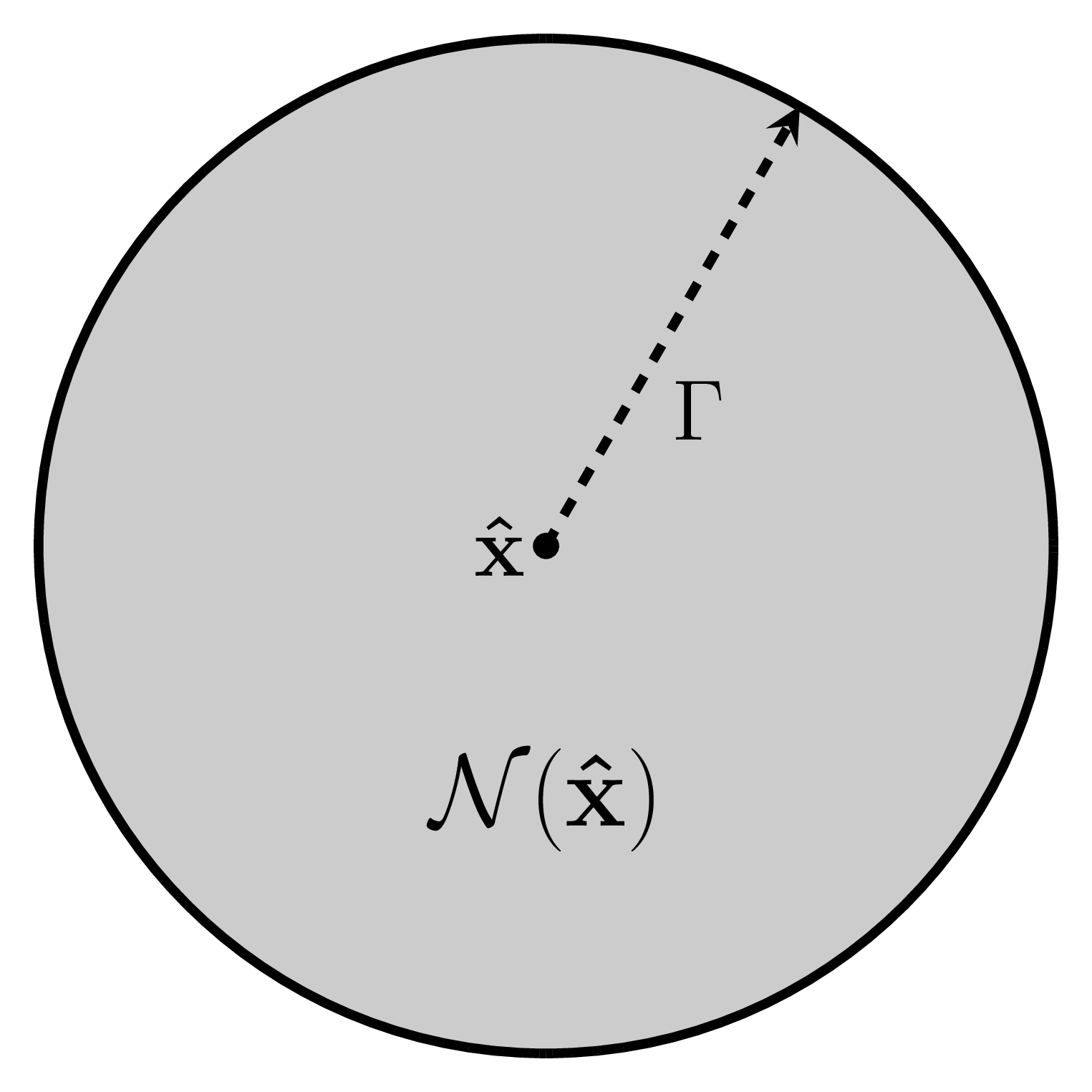}
         \caption{}
         \label{fig:neighbourhood_of_xhat}
     \end{subfigure}
     \begin{subfigure}[b]{0.5\textwidth}
         \centering
         \includegraphics[width=\textwidth]{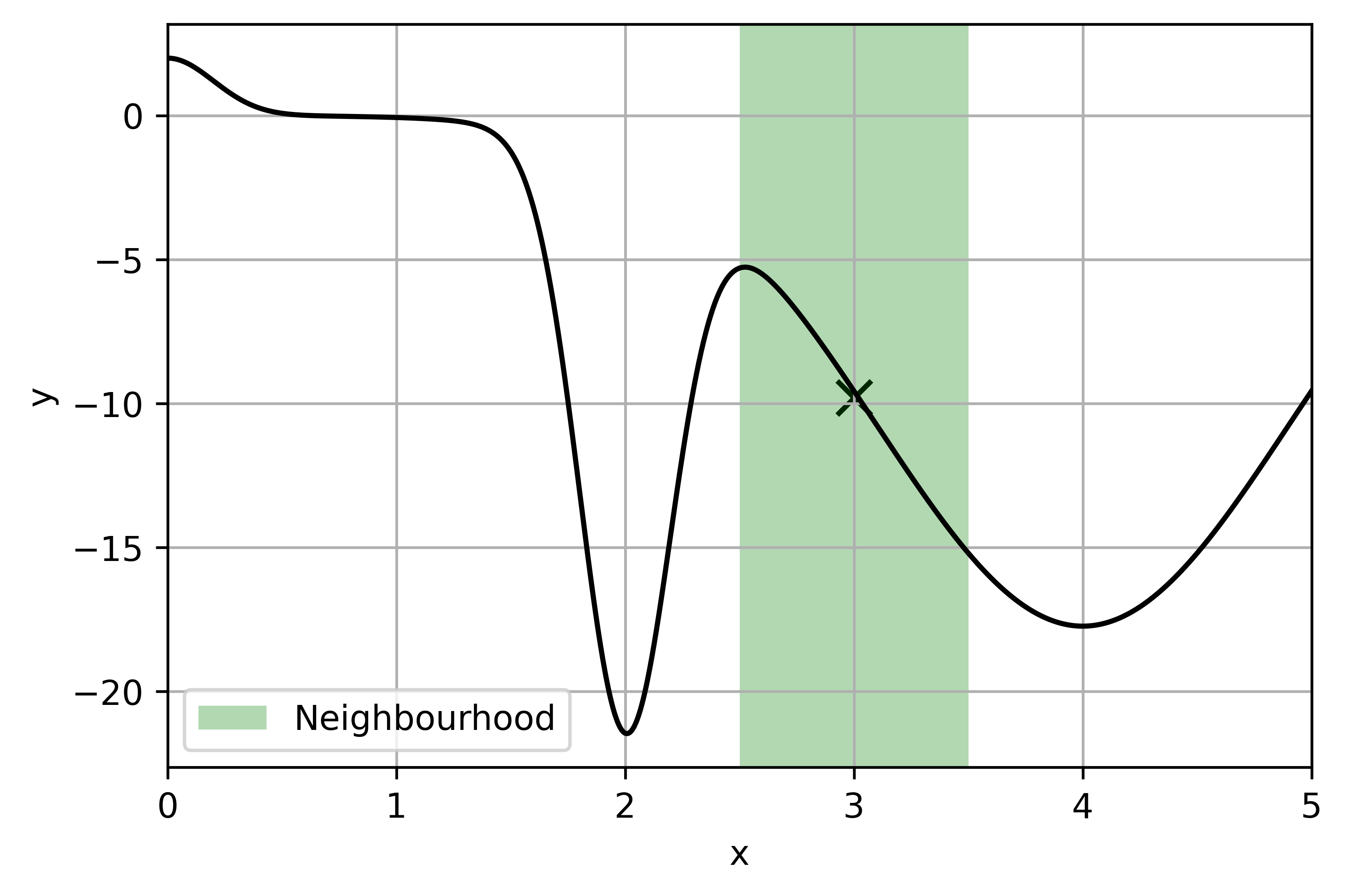}
         \caption{}
         \label{fig:neighbourhood_of_xhat_1d_eg}
     \end{subfigure}
        \caption{(a): A depiction of the neighbourhood of a point $\mathbf{\hat{x}}$. (b): Neighbourhood (green hue) of a point, depicted as a black cross, for the simple one dimensional example from section \ref{constrained:prob_def}.}
        \label{fig:neighbourhoods}
\end{figure}

An example of a point $\mathbf{\hat{x}}$ and its neighbourhood is given in Figure \ref{fig:neighbourhood_of_xhat}. To accompany this abstract figure, we also present Figure \ref{fig:neighbourhood_of_xhat_1d_eg} which uses the same one dimensional example used in section \ref{constrained:prob_def}. Here we see a point depicted as a black cross and its neighbourhood ($\Gamma = 0.5$), depicted in a green hue.

\begin{figure}
     \centering
     \begin{subfigure}[b]{0.45\textwidth}
         \centering
         \includegraphics[width=\textwidth]{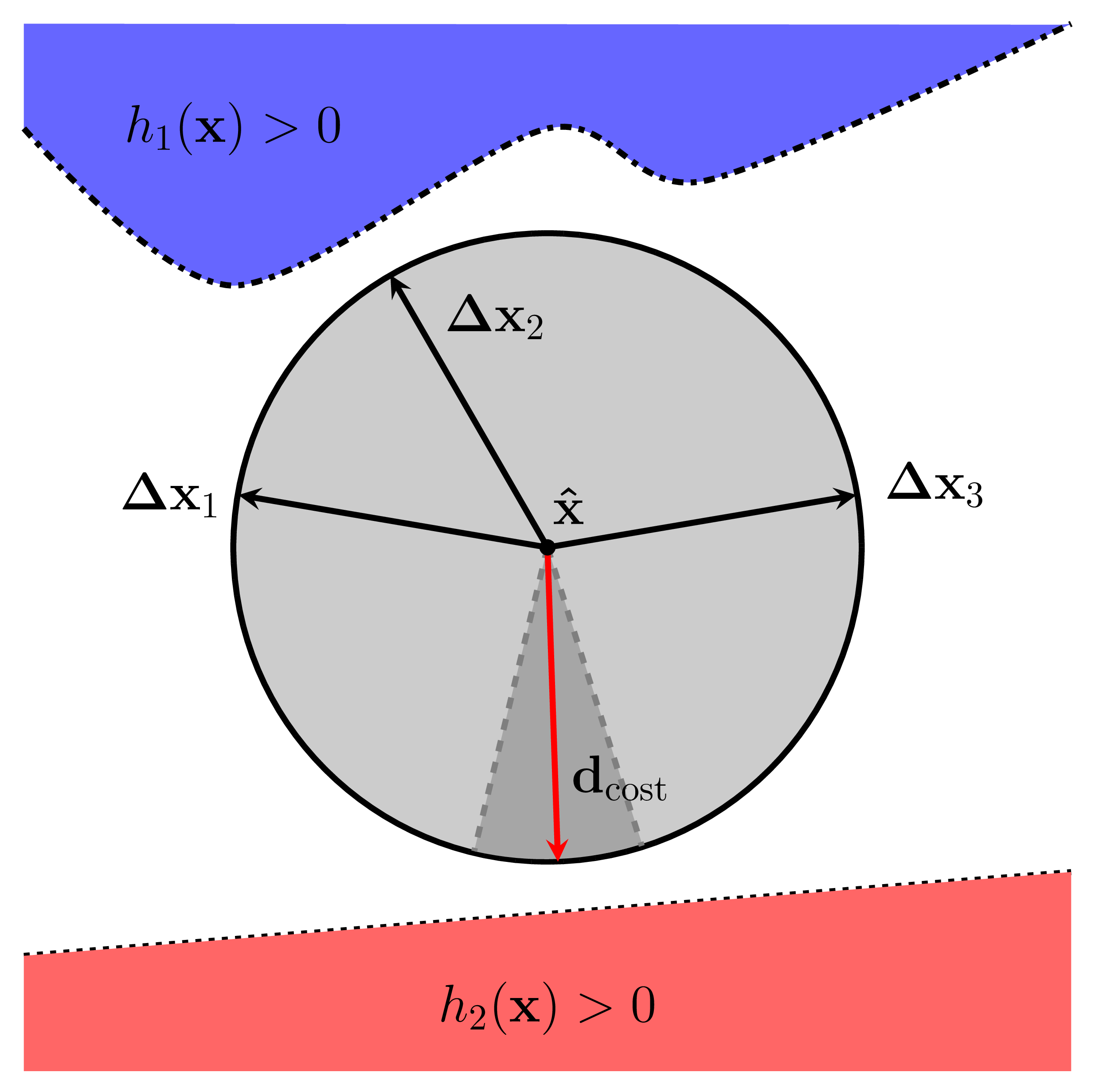}
         \caption{}
         \label{fig:no_constraint_vio}
     \end{subfigure}
     \hfill
     \begin{subfigure}[b]{0.45\textwidth}
         \centering
         \includegraphics[width=\textwidth]{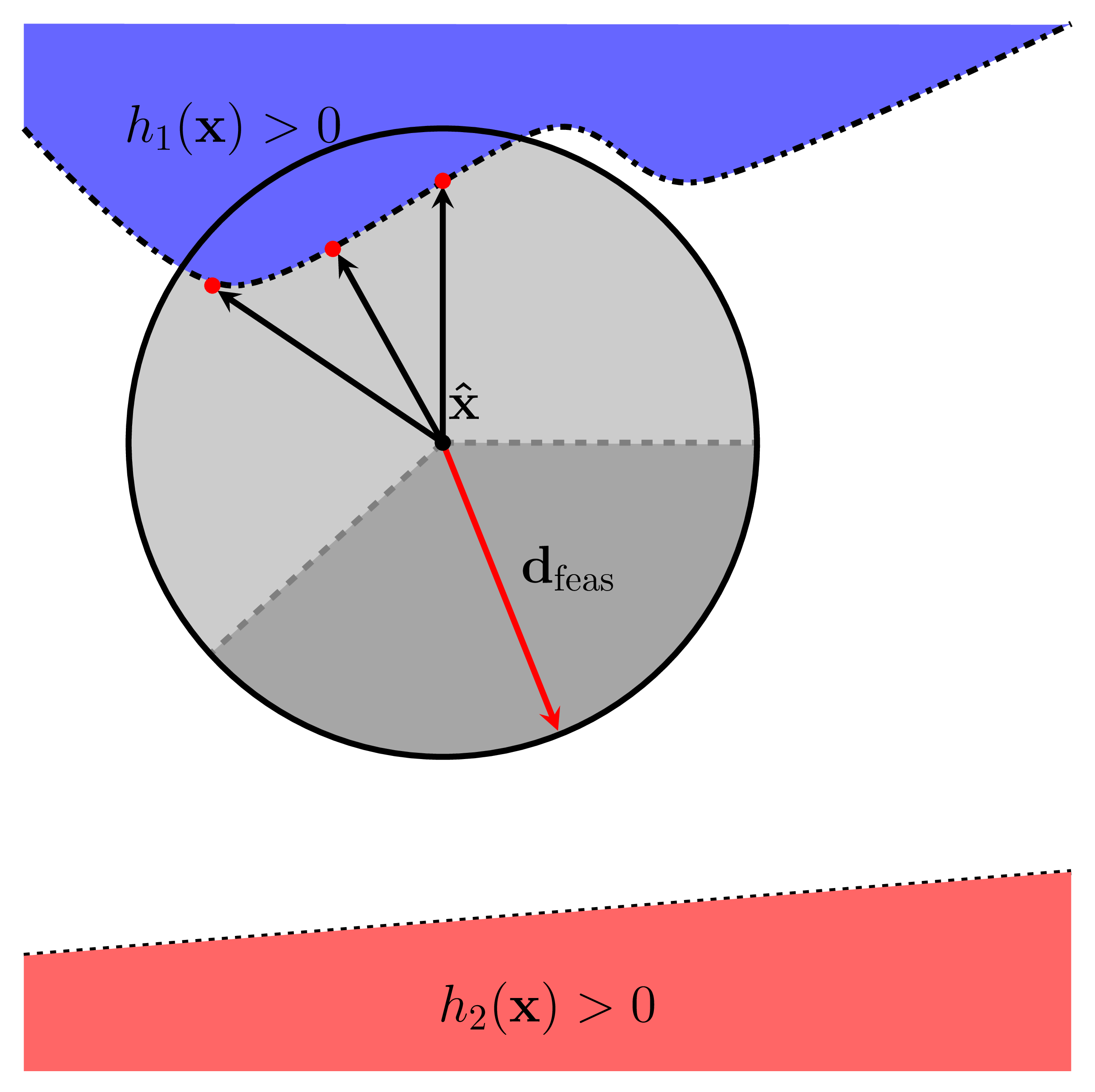}
         \caption{}
         \label{fig:constraint_vio}
     \end{subfigure}
        \caption{Both figures depict infeasible regions in blue and red. (a): There are no constraint violations in the neighbourhood of the point i.e. it is feasible under perturbations. The algorithm's goal is to reduce the worst-case cost. (b): The neighbourhood of point $\mathbf{\hat{x}}$ intersects with an infeasible region, rendering the point infeasible under perturbations. The algorithm's goal is to find a descent direction to guide the iterate back into the feasible region.}
        \label{fig:arrtoc_constraints}
\end{figure}

Given this point, \textbf{Step 1} of the algorithm is to conduct an exploration of the neighbourhood for high-cost directions by solving the inner maximization problem, $\underset{\boldsymbol{\Delta}\mathbf{x} \in \mathcal{U}}{\text{max}} \; f(\mathbf{\hat{x}} + \boldsymbol{\Delta}\mathbf{x})$ via a multi-start gradient ascent algorithm. The details of this will be explored in section \ref{constrained:step_1}. \textbf{Step 2} also involves an exploration of the neighbourhood, but for constraint-violating directions, by solving the constraint maximization problem, $\underset{\boldsymbol{\Delta}\mathbf{x} \in \mathcal{U}}{\text{max}} \; h_j(\mathbf{\hat{x}} + \boldsymbol{\Delta}\mathbf{x})$. This is also done via a multi-start gradient ascent algorithm, the details of which will be explored in section \ref{constrained:step_2}. Following these two steps, two outcomes are possible as depicted in Figure \ref{fig:arrtoc_constraints}. In the first case, Figure \ref{fig:no_constraint_vio} shows an iterate, where during \textbf{Step 2} of the algorithm, no constraint violating neighbours were found in the neighbourhood of the point i.e. it is feasible under perturbations. In this scenario, cost considerations take precedence over constraints. The objective is to identify a robust local move that reduces the worst-case cost. This involves finding a direction, $\mathbf{d}_{\text{cost}}$, and step-size, $\rho_{\text{cost}}$, which moves away from the high-cost directions, depicted as black arrows, which were found during \textbf{Step 1} of the algorithm. The direction and step-size are found by solving a Second Order Cone Program (SOCP) which will be detailed in section \ref{constrained:step_3a}. On the other hand, Figure \ref{fig:constraint_vio} shows an iterate, where during \textbf{Step 2} of the algorithm, constraint violating neighbours were found, rendering it infeasible under perturbations. In this scenario, constraints take precedence over cost, with the goal being to find a robust local move which guides the iterate back into the feasible region. Again, we must find a direction, $\mathbf{d}_{\text{feas}}$, and step-size, $\rho_{\text{feas}}$, which moves away from the constraint-violating neighbour directions, depicted as black arrows. These are also found by solving a SOCP which will be detailed in section \ref{constrained:step_3b}. The algorithm iteratively follows the above steps until it reaches a robust local minimum, where (i) the point is surrounded by high-cost neighbours on all sides, meaning there is no direction to reduce the worst-case cost, and (ii) the neighbourhood of the point does not intersect with any of the constraint-violating regions.

A high-level psuedocode of the main steps of the ARRTOC algorithm is provided in Pseudocode \ref{arrtoc_main_steps_algo}. In the following sections, we consider each of the steps in Pseudocode \ref{arrtoc_main_steps_algo} in detail in order to make our intuitive understanding more rigorous. 

\begin{algorithm}
\caption{The Main Steps of the ARRTOC Algorithm}\label{arrtoc_main_steps_algo}
\begin{algorithmic}[1]
\State \textbf{Step 1:} \textit{Neighbourhood Cost Exploration} - \textbf{See section \ref{constrained:step_1}}.
\State \textbf{Step 2:} \textit{Neighbourhood Constraint Exploration} - \textbf{See section \ref{constrained:step_2}}.
\State \textbf{Step 3A:} \textit{Robust Local Move If Feasible Under Perturbations} - \textbf{See section \ref{constrained:step_3a}}.
\State \textbf{Step 3B:} \textit{Robust Local Move If Infeasible Under Perturbations} - \textbf{See section \ref{constrained:step_3b}}.
\end{algorithmic}
\end{algorithm}

\subsection{Step 1: Neighbourhood Cost Exploration}\label{constrained:step_1}

As discussed in section \ref{constrained:intuition}, the neighbourhood of a point $\mathbf{\hat{x}}$, must be explored to identify the set of high cost directions (i.e. the set of all worst possible implementation errors) at this point:

\begin{equation}\label{eq:unconstrained_worst_possible_neighbours}
\mathcal{U}^*(\mathbf{\hat{x}}) = \left\{\boldsymbol{\Delta}\mathbf{x}^* \;| 
 \; \text{arg}\underset{\boldsymbol{\Delta}\mathbf{x} \in \mathcal{U}}{\text{max}} \; f(\mathbf{\hat{x}} + \boldsymbol{\Delta}\mathbf{x}) \right\}
\end{equation}
This requires the identification of all of the global maxima of the inner maximization problem, $\underset{\boldsymbol{\Delta}\mathbf{x} \in \mathcal{U}}{\text{max}} \; f(\mathbf{\hat{x}} + \boldsymbol{\Delta}\mathbf{x})$. However, globally solving the inner maximization problem is not trivial for non-convex problems \cite{Bertsimas2010}. To avoid this, a heuristic neighbourhood exploration algorithm can be adopted which relies on a "thorough" exploration of the neighbourhood of $\mathbf{\hat{x}}$.

\begin{figure}
\begin{center}
\includegraphics[width=0.35\textwidth]{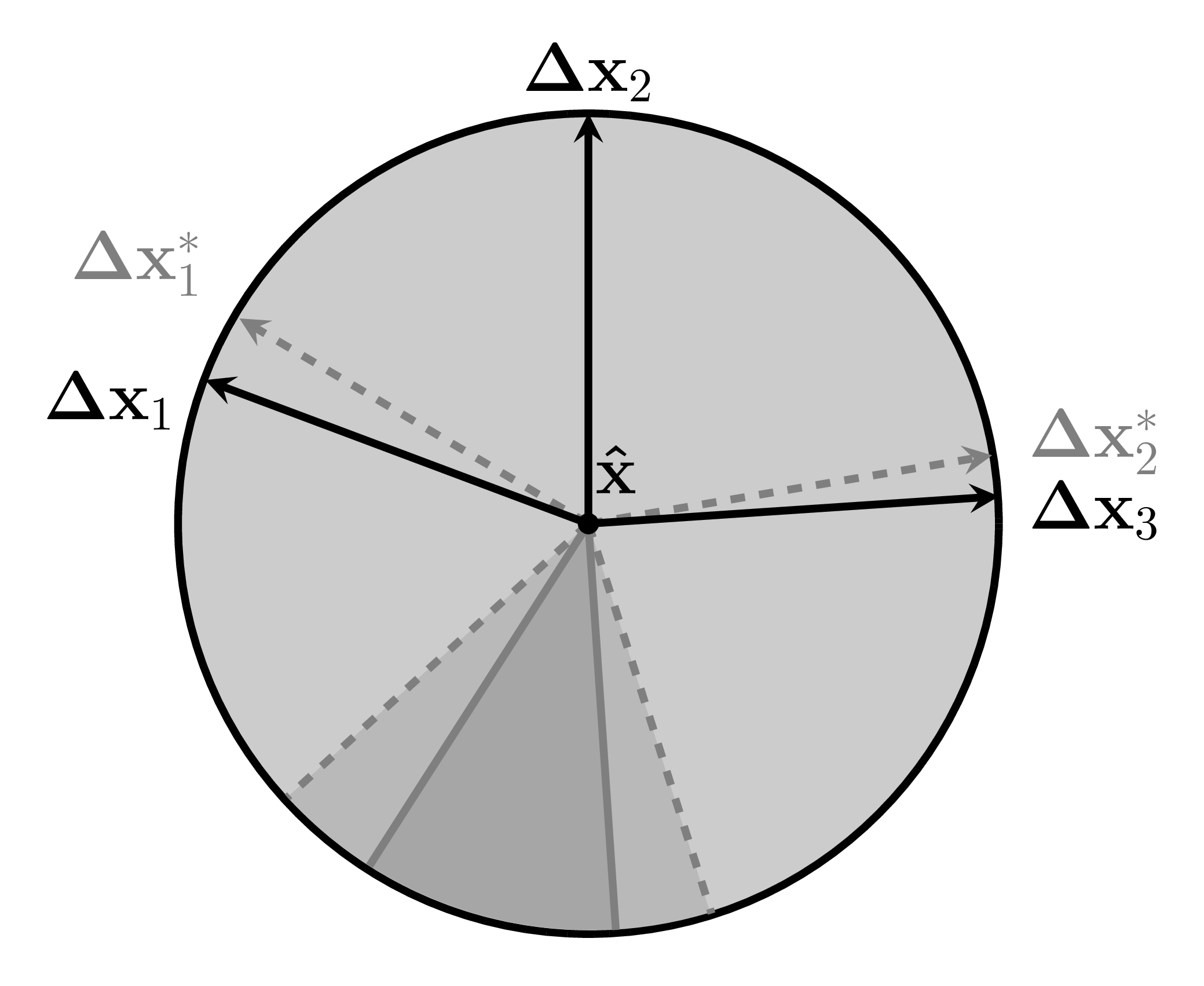} 
\caption{A successful application of a neighbourhood exploration. Global maximizers, from Eq. (\ref{eq:unconstrained_worst_possible_neighbours}), are indicated by light grey dashed arrows. Descent directions based on these maximizers is shown as a light grey cone with dashed lines. No knowledge of these is assumed. Instead, the inner-maximization problem is solved to give local solutions, represented as black arrows. Crucially, the set of descent directions derived from the local maximizers (dark grey cone with solid lines) is a subset of the desired descent directions based on the global maximizers.}
\label{fig:local_search_successful}
\end{center}
\end{figure}

To understand how the heuristic neighbourhood exploration operates, consider the scenario depicted in Figure \ref{fig:local_search_successful} which shows a point, $\mathbf{\hat{x}}$ and its neighbourhood. Within this neighbourhood, we depict the worst implementation errors (from equation (\ref{eq:unconstrained_worst_possible_neighbours})), represented by light grey dashed arrows. As noted, these may be impractical to identify. Instead, we approximately solve the inner maximization problem, $\underset{\boldsymbol{\Delta}\mathbf{x} \in \mathcal{U}}{\text{max}} \; f(\mathbf{\hat{x}} + \boldsymbol{\Delta}\mathbf{x})$, to find local maximizers. We collect these local solutions in a set $\mathcal{M}$, denoted as black arrows in Figure \ref{fig:local_search_successful}. Notably, the local solutions in $\mathcal{M}$ define a cone that encompasses all global maximizers, $\boldsymbol{\Delta}\mathbf{x}^*_i \in \mathcal{U}^*(\mathbf{\hat{x}})$. This implies:

\begin{equation}\label{eq:local_minimizer_condition}
\begin{aligned}
&\boldsymbol{\Delta}\mathbf{x}^* =\sum_{\boldsymbol{\Delta}\mathbf{x}_i \in \mathcal{M}} \alpha_i \boldsymbol{\Delta}\mathbf{x}_i \\ &\forall \; \boldsymbol{\Delta}\mathbf{x}^* \in \mathcal{U}^*(\mathbf{\hat{x}}) 
\end{aligned}
\end{equation}
where $\alpha_i \in \mathbb{R} \geq 0$. This geometrically corresponds to the condition that the global maximizers are contained within the cone defined by the local maximizers as per Figure \ref{fig:local_search_successful}.

The set of possible descent directions based on the global maximizers (which we have no knowledge of) is defined as those vectors forming angles exceeding 90 degrees with all of the worst implementation errors, as defined by equation (\ref{eq:unconstrained_worst_possible_neighbours}):

\begin{equation}\label{eq:descent_direction_condition}
\begin{aligned}
&\mathbf{d}\boldsymbol{\Delta}\mathbf{x}^* < 0  \\ &\forall \; \boldsymbol{\Delta}\mathbf{x}^* \in \mathcal{U}^*(\mathbf{\hat{x}}) 
\end{aligned}
\end{equation}
where $\mathbf{d} \in \mathbb{R}^{n_x}$, represents a potential descent direction. These descent directions are depicted as a light shade of grey demarcated by dashed grey lines. Similarly, the set of descent directions based on the local maximizers i.e. those which satisfy:
\begin{equation}\label{eq:descent_direction_local_condition}
\begin{aligned}
&\mathbf{d}\boldsymbol{\Delta}\mathbf{x} < 0  \\ &\forall \; \boldsymbol{\Delta}\mathbf{x} \in \mathcal{M}
\end{aligned}
\end{equation}
is depicted as a dark shade of grey demarcated by solid grey lines. Critically, notice that the set of descent directions based on the local maximizers (dark grey shaded cone) is a subset of the set of descent directions based on the global maximzers (light grey shaded cone). Hence, by satisfying equation (\ref{eq:local_minimizer_condition}), we ensure that our set of descent directions is a subset of the desired directions obtained from the global maximizers. Failing to do this, could lead to a set of descent directions which inadvertently increases, rather than decreases, the worst-case cost. 

Therefore, a "thorough" neighbourhood exploration algorithm should prioritise the discovery of local solutions that satisfy equation (\ref{eq:local_minimizer_condition}). To achieve this, a heuristic multi-start neighbourhood exploration algorithm is proposed. Although this algorithm does not offer a guarantee for satisfying equation (\ref{eq:local_minimizer_condition}), its semi-exhaustive nature provides confidence in the obtained result \cite{Bertsimas2010}. The details of this algorithm are outlined in the next section.

\subsubsection{A Heuristic Multi-Start Neighbourhood Exploration Algorithm}\label{unconstrained:heuristic_neighbourhood}

The multi-start gradient ascent algorithm which is used to approximately solve the inner maximization problem utilises $n_x + 1$ gradient ascents by starting at different points within the neighbourhood of the current iterate, $\mathbf{x}^k$. The algorithm is defined in Pseudocode \ref{exploration_algo}.

\begin{algorithm}
\caption{Heuristic Multi-Start Neighbourhood Exploration Algorithm}\label{exploration_algo}
\begin{algorithmic}[1]
\Require Current Iterate, $\mathbf{x}^k$; History Set From Previous Iteration, $\mathcal{H}^{k-1}$; Neighbourhood Upper Bound, $\Gamma$
\Ensure Newly Recorded History Set, $\mathcal{H}^{k}$
\State $\mathcal{H}^{k} \gets \mathcal{H}^{k-1}$
\For{$i \gets 1$ to $n_x + 1$} 
\If{$i = 1$}
\State $\boldsymbol{\Delta}\mathbf{x}_{\text{INIT}} = \mathbf{0}$
\Else 
\State $\boldsymbol{\Delta}\mathbf{x}_{\text{INIT}} = sign\left(\frac{\partial f(\mathbf{x} = \mathbf{x}^k)}{\partial x_i}\right)\frac{\Gamma}{3}\mathbf{e}_i$ where $\mathbf{e}_i$ is the unit-vector along the $i$th coordinate.
\EndIf
\State Solve $\underset{\boldsymbol{\Delta}\mathbf{x} \in \mathcal{U}}{\text{max}} \; f(\mathbf{x}^k + \boldsymbol{\Delta}\mathbf{x})$ using diminishing step size gradient ascent algorithm, given initial point $\boldsymbol{\Delta}\mathbf{x}_{\text{INIT}}$,\newline\hspace*{1.5em}initial step size, $\frac{\Gamma}{5}$, which is reduced by $0.99$ after every step, and record all function evaluations $\left(\mathbf{x}, f(\mathbf{x})\right)$ in\newline\hspace*{1.5em}history set $\mathcal{H}^{k}$
\EndFor
\State
\Return $\mathcal{H}^{k}$
\end{algorithmic}
\end{algorithm}
The outcome of the neighbourhood exploration algorithm is the history set, $\mathcal{H}^{k}$. This set comprises all the function evaluations, $\left(\mathbf{x}, f(\mathbf{x})\right)$, made during the multiple gradient ascents in the $k$th iteration of the ARRTOC algorithm. The history set allows for the construction of the set of worst implementation errors or neighbours with high cost, denoted as $\mathcal{M}^k$ i.e. the local maximizers of the inner maximization problem. Subsequently, if the iterate is found to be feasible under perturbations, this set is used to determine the descent direction as will be detailed in section \ref{constrained:step_3a}. Due to the multi-start nature of the algorithm, a comprehensive neighbourhood exploration is conducted to heuristically satisfy the condition outlined in equation (\ref{eq:local_minimizer_condition}), as depicted in Figure \ref{fig:local_search_successful}.

\subsection{Step 2: Neighbourhood Constraint Exploration}\label{constrained:step_2}

In most ARO algorithms, only a neighbourhood cost exploration is conducted as constraints do not play an active role \cite{Bertsimas2010, Bogunovic2018, Paulson2020}. However, when constraints are included, it is also important to determine if there are any neighbours that violate these constraints to determine whether the current iterate is feasible under perturbations. This step crucially dictates whether \textbf{Step 3A} (section \ref{constrained:step_3a}) or \textbf{Step 3B} (section \ref{constrained:step_3b}) is conducted after the neighbourhood explorations. 

The neighbourhood constraint exploration is conducted by solving the constraint maximization problem for all constraints in the set of constraints, $\mathcal{J}$:

\begin{equation}\label{eq:constraint_max}
\underset{\boldsymbol{\Delta}\mathbf{x} \in \mathcal{U}}{\text{max}} \; h_j(\mathbf{x} + \boldsymbol{\Delta}\mathbf{x})  \;\;\;\; \forall j \in \mathcal{J}
\end{equation}

Readers should observe the resemblance between solving the inner maximization problem, $\underset{\boldsymbol{\Delta}\mathbf{x} \in \mathcal{U}}{\text{max}} \; f(\mathbf{x} + \boldsymbol{\Delta}\mathbf{x})$, and the constraint maximization problem of equation (\ref{eq:constraint_max}). Due to the fact that they are almost identical problems, we repurpose the Heuristic Multi-Start Neighbourhood Exploration algorithm in Pseudocode \ref{exploration_algo}, for the constraint maximization problem of equation (\ref{eq:constraint_max}). Specifically, in \textbf{Step 8} of Pseudocode \ref{exploration_algo}, instead of solving the inner maximization problem we simply solve equation (\ref{eq:constraint_max}) $\forall j \in \mathcal{J}$. Moreover, instead of recording all function evaluations in a history set, we instead record any neighbours, $\mathbf{x}$, that have a constraint value exceeding zero (i.e. a constraint-violating neighbour) in a constraint history set, $\mathcal{C}^k$. 

By the completion of step 1 and step 2 in the ARRTOC algorithm, we obtain two important sets. Firstly, from step 1, we have the history set $\mathcal{H}^{k}$, which is used to construct the set of neighbours with high cost, denoted as $\mathcal{M}^{k}$. Secondly, from step 2, we have the constraint history set $\mathcal{C}^k$, which allows us to evaluate the feasibility of the current iterate $\mathbf{x}^k$ under perturbations. If the iterate is feasible, then cost takes precedence over constraints and \textbf{Step 3A} as detailed in section \ref{constrained:step_3a} is performed, otherwise, \textbf{Step 3B} as detailed in section \ref{constrained:step_3b} is performed.

\subsection{Step 3A: Robust Local Move If Feasible Under Perturbations}\label{constrained:step_3a}

If the current iterate, $\mathbf{x}^k$, has no neighbours in the constraint history set, $\mathcal{C}^k$, then it is feasible under perturbations. This is equivalent to the scenario depicted in Figure \ref{fig:no_constraint_vio}. Therefore, we want to find a direction, $\mathbf{d}_{\text{cost}}$, and step-size, $\rho_{\text{cost}}$, which moves in the opposite direction to all of the worst-case implementation errors (or high-cost neighbours) found during \textbf{Step 1} of the algorithm (section \ref{constrained:step_1}). In this section, we explore this in detail.

\subsubsection{Finding a Descent Direction}\label{constrained:step_3a_direction}

The first goal of finding a possible descent direction is to construct set $\mathcal{M}^k$, the set of worst implementation errors or neighbours with high cost of the current iterate, from the history set $\mathcal{H}^{k}$. This is defined as follows:

\begin{equation}\label{eq:worst_neighbours_current_iterate}
\mathcal{M}^k = \left\{\mathbf{x} \;| \mathbf{x} \in \mathcal{H}^{k}; \mathbf{x} \in \mathcal{N}^{k};  f(\mathbf{x}) \geq \Tilde{g}(\mathbf{x}^k) - \sigma^k\right\}
\end{equation}
where $\mathcal{H}^{k}$ is the history set, $\mathcal{N}^{k}$ is the neighbourhood of the current iterate as defined in equation (\ref{eq:unconstrained_neighbourhood_set}), $\Tilde{g}(\mathbf{x}^k)$ is the estimated worst case cost of the current iterate i.e. the neighbour of the current iterate which has the largest cost found during the neighbourhood exploration. Finally, and most importantly, $\sigma^k$ is a cost factor which dictates the size of $\mathcal{M}^k$ by reducing the threshold of entry to the set, in order to ensure a feasible move is initially found. $\sigma^k$ plays a crucial role in verifying whether the current iterate is a robust local minimum in case no descent directions are found. This will be detailed in the next section, section \ref{constrained:step_3a_verify}. For the first iteration, $\sigma^1 = 0.2 \left(\Tilde{g}(\mathbf{x}^1) - f(\mathbf{x}^1)\right)$ and $\sigma^k = \sigma^{k-1}$ in all iterations thereafter unless a robust local minimum needs to be verified. 

Given all of the bad neighbours of the current iterate in $\mathcal{M}^k$, we wish to find a descent direction $\mathbf{d}_{\text{cost}}$ which points away from all of the worst implementation errors. In other words, we seek the descent direction which forms the largest angle with each of the worst implementation errors. Importantly, this descent direction must also satisfy equation (\ref{eq:descent_direction_local_condition}). This can be formulated as a constrained optimization problem, specifically a Second Order Cone Program (SOCP):

\begin{equation}\label{eq:socp}
\begin{aligned}
\underset{\mathbf{d}_{\text{cost}}, \beta}{\text{min}} \; &\beta \\
\text{s.t.} \; &\left||\mathbf{d}_{\text{cost}}\right|| \leq 1 \\
& \mathbf{d}_{\text{cost}}\left(\frac{\mathbf{x}_i - \mathbf{x}^k}{\left||\mathbf{x}_i - \mathbf{x}^k\right||}\right) \leq \beta \;\;\;\; \forall \mathbf{x}_i \in \mathcal{M}^k \\
&\beta \leq -\epsilon
\end{aligned}
\end{equation}
where $\epsilon$ is a small positive scalar which we set to $0.01$ in our case studies in section \ref{sec:results_discussion}. It is worth noting that the expression $\mathbf{x}_i - \mathbf{x}^k = \boldsymbol{\Delta}\mathbf{x} \in \mathcal{U}$ defines the implementation errors, where the second constraint in equation (\ref{eq:socp}) corresponds to equation (\ref{eq:descent_direction_local_condition}), as intended.

Upon solving equation (\ref{eq:socp}), if the problem is infeasible then this implies that no descent direction exists which forms a 90 degree angle or greater with all of the worst implementation errors. This suggests that the current iterate is surrounded by high-cost neighbours on all sides, and it may be a robust local minimum. The verification of this scenario is discussed in section \ref{constrained:step_3a_verify}).

On the other hand, if the problem is feasible, it indicates the existence of a descent direction. The subsequent objective is to determine an appropriate step size along this direction. This aspect is addressed in section \ref{constrained:step_3a_step_size}).

\subsubsection{Verifying a Robust Local Minimum}\label{constrained:step_3a_verify}

If equation (\ref{eq:socp}) is infeasible, then this implies that the current iterate, $\mathbf{x}^k$, is surrounded by bad neighbours i.e. no direction in the neighbourhood exists which reduces the worst-case cost. This point may be a robust local minimum which needs to be verified. In particular, it may be that in constructing set $\mathcal{M}^k$, as per equation (\ref{eq:worst_neighbours_current_iterate}), the value of the cost factor, $\sigma^k$, was too large and so we were too lenient in our definition of a bad neighbour. In order to verify this, we iteratively reduce the value of $\sigma^k$ by a factor of $1.05$ and re-solve equation (\ref{eq:socp}) until it is feasible and a descent direction is found. If however, $\sigma^k$ drops below a threshold value, $\sigma_{\text{MIN}}$, which we choose to be $0.001$ in our case studies in section \ref{sec:results_discussion}, then the algorithm terminates and $\mathbf{x}^k$ is defined as the robust local minimum. 

\subsubsection{Finding the Step Size}\label{constrained:step_3a_step_size}

If equation (\ref{eq:socp}), is feasible then a descent direction exists as per the red arrow in Figure \ref{fig:no_constraint_vio}. The next goal is to find an appropriate step-size, $\rho_{\text{cost}}$, along this direction. In particular, the step size should be sufficiently large so that the bad neighbours of the current iterate, represented by the set $\mathcal{M}^k$, are excluded from the neighbourhood of the new iterate $\mathbf{x}^{k+1} = \mathbf{x}^{k} + \rho_{\text{cost}}^*\mathbf{d}_{\text{cost}}^*$. This ensures that, at the minimum, these bad neighbours lie on the boundary of the new neighbourhood. This approach helps avoid a redundant neighbourhood exploration for the new iterate. Given, $\mathbf{x}^{k+1} = \mathbf{x}^{k} + \rho_{\text{cost}}^*\mathbf{d}_{\text{cost}}^*$, then the distance, $\left|\left|\mathbf{x}_i - \mathbf{x}^{k+1}\right|\right|, \; \forall \mathbf{x}_i \in \mathcal{M}^k$, can be easily calculated via the cosine rule:

\begin{equation}\label{eq:distance}
\left|\left|\mathbf{x}_i - \mathbf{x}^{k+1}\right|\right|^2 = \rho_{\text{cost}}^2 + \left|\left|\mathbf{x}_i - \mathbf{x}^{k}\right|\right|^2 - 2\rho_{\text{cost}}\left(\mathbf{x}_i - \mathbf{x}^{k}\right)\mathbf{d}_{\text{cost}}^* \;\;\;\; \forall \mathbf{x}_i \in \mathcal{M}^k
\end{equation}
Hence, we can find the optimal step size that positions the bad neighbours of the current iterate at the minimum on the boundary of the new neighbourhood as an optimization problem:

\begin{equation}\label{eq:distance_optimization}
\begin{aligned}
\rho_{\text{cost}}^* = \text{arg }\underset{\rho_{\text{cost}}}{\text{min}} & \; \rho_{\text{cost}} \\
\text{s.t.} & \; \rho_{\text{cost}} \geq \left(\mathbf{x}_i - \mathbf{x}^{k}\right)\mathbf{d}_{\text{cost}}^* + \sqrt{\left(\left(\mathbf{x}_i - \mathbf{x}^{k}\right)\mathbf{d}_{\text{cost}}^*\right)^2 - \left|\left|\mathbf{x}_i - \mathbf{x}^{k}\right|\right|^2 + \Gamma^2} \;\;\;\; \forall \mathbf{x}_i \in \mathcal{M}^k
\end{aligned}
\end{equation}
where $\rho_{\text{cost}}^*$ is the optimal step-size. This problem does not require a formal algorithm, instead it can be solved with $\left|\mathcal{M}^k\right|$ iterations by checking the constraint in equation (\ref{eq:distance_optimization}) numerically and setting $\rho_{\text{cost}}^*$ to the minimum value found.

\subsubsection{Robust Local Move Algorithm}\label{constrained:step_3a_final_algo}

Finally, putting everything for step 3A together, the robust local move algorithm is detailed in Psuedocode \ref{robust_local_move_algo}.

\begin{algorithm}
\caption{Robust Local Move Algorithm}\label{robust_local_move_algo}
\begin{algorithmic}[1]
\Require Current Iterate, $\mathbf{x}^k$; History Set From Neighbourhood Cost Exploration, $\mathcal{H}^{k}$; Cost Factor From Previous Iteration $\sigma^{k-1}$; Cost Factor Threshold $\sigma_{\text{MIN}}$
\Ensure Robust Local Minimum, $\mathbf{x}^*$ or Optimal Step-Size and Descent Direction $\left(\rho_{\text{cost}}^*, \mathbf{d}_{\text{cost}}^*\right)$
\State $\sigma^{k} \gets \sigma^{k-1}$
\While{$\sigma^{k} > \sigma_{\text{MIN}}$}
\State Construct set $\mathcal{M}^k$ as per equation (\ref{eq:worst_neighbours_current_iterate})
\State Solve SOCP - Equation (\ref{eq:socp})
\If{Equation (\ref{eq:socp}) is infeasible}
\State $\sigma^{k} \gets \frac{\sigma^{k}}{1.05}$
\Else
\State Find optimal step-size $\rho_{\text{cost}}^*$ by solving equation (\ref{eq:distance_optimization})
\State
\Return $\left(\rho_{\text{cost}}^*, \mathbf{d}_{\text{cost}}^*\right)$
\EndIf
\EndWhile
\State $\mathbf{x}^* \gets \mathbf{x}^k$, Equation (\ref{eq:socp}) (SOCP) remained infeasible and so $\mathbf{x}^k$, was verified as the robust local minimum
\State
\Return $\mathbf{x}^*$
\end{algorithmic}
\end{algorithm}

\subsection{Step 3B: Robust Local Move If Infeasible Under Perturbations}\label{constrained:step_3b}
 
If the current iterate, $\mathbf{x}^k$, has neighbours in the constraint history set, $\mathcal{C}^k$, then it is infeasible under perturbations. This is equivalent to the scenario depicted in Figure \ref{fig:constraint_vio}. Therefore, we want to find a direction, $\mathbf{d}_{\text{feas}}$, which moves in the opposite direction from all of the constraint violating neighbours and a step-size, $\rho_{\text{feas}}$, which places these neighbours, at the minimum, on the boundary of the new neighbourhood. 

Once again, readers should note the resemblance of this problem to that of section \ref{constrained:step_3a} of finding a robust local move if feasible under perturbations. However, in this case, instead of worst-cost neighbours, we consider constraint-violating neighbours, otherwise the problem is identical. Consequently, we repurpose the approach adopted in section \ref{constrained:step_3a}. Specifically, the feasible descent direction is found by solving a similar SOCP:

\begin{equation}\label{eq:socp_constraint}
\begin{aligned}
\underset{\mathbf{d}_{\text{feas}}, \beta}{\text{min}} \; &\beta \\
\text{s.t.} \; &\left||\mathbf{d}_{\text{feas}}\right|| \leq 1 \\
& \mathbf{d}_{\text{feas}}\left(\frac{\mathbf{x}_i - \mathbf{x}^k}{\left||\mathbf{x}_i - \mathbf{x}^k\right||}\right) \leq \beta \;\;\;\; \forall \mathbf{x}_i \in \mathcal{C}^k \\
&\beta \leq -\epsilon
\end{aligned}
\end{equation}
As depicted in Figure \ref{fig:constraint_vio}, the optimal feasible descent direction, $\mathbf{d}_{\text{feas}}^*$, creates the largest angle with all of the implementation errors which lead to constraint violations (i.e. $\mathbf{x}_i - \mathbf{x}^k, \forall \mathbf{x}_i \in \mathcal{C}^k$).

The step-size to take along the feasible descent direction is also found in a similar manner as detailed in section \ref{constrained:step_3a} by solving equation (\ref{eq:distance_optimization}) but for all $\mathbf{x}_i$ in the constraint history set, $\mathcal{C}^k$, as opposed to the set $\mathcal{M}^k$:

\begin{equation}\label{eq:distance_optimization_constraint}
\begin{aligned}
\rho_{\text{feas}}^* = \text{arg }\underset{\rho_{\text{feas}}}{\text{min}} & \; \rho_{\text{feas}} \\
\text{s.t.} & \; \rho_{\text{feas}} \geq \left(\mathbf{x}_i - \mathbf{x}^{k}\right)\mathbf{d}_{\text{feas}}^* + \sqrt{\left(\left(\mathbf{x}_i - \mathbf{x}^{k}\right)\mathbf{d}_{\text{feas}}^*\right)^2 - \left|\left|\mathbf{x}_i - \mathbf{x}^{k}\right|\right|^2 + \Gamma^2} \;\;\;\; \forall \mathbf{x}_i \in \mathcal{C}^k
\end{aligned}
\end{equation}
Following the solution of equations (\ref{eq:socp_constraint}) and (\ref{eq:distance_optimization_constraint}), both the optimal feasible step-size, $\rho_{\text{feas}}^*$, and descent direction, $\mathbf{d}_{\text{feas}}^*$, should be known. These can be used to give the new iterate, $\mathbf{x}^{k+1} = \mathbf{x}^{k} + \rho_{\text{feas}}^*\mathbf{d}_{\text{feas}}^*$, which moves away from the infeasible region and towards the feasible region.

\section{Results and Discussion}\label{sec:results_discussion}

This section presents the application of the ARRTOC algorithm to three progressively complex case studies. The first case study in section \ref{rd:illustrative_example}, serves as an illustrative example, visually demonstrating the fundamental principles and insights underlying the ARRTOC algorithm. Subsequently, we leverage these insights in the practical case studies presented in section \ref{rd:bioreactor} and \ref{rd:evaporator}.

\subsection{Illustrative Example}\label{rd:illustrative_example}

In this section, our aim is to explore the core concepts of the ARRTOC algorithm through an illustrative example. We focus here on the RTO problem without considering the underlying control problem explicitly, although we do consider hypothetical control scenarios to highlight important findings. The nominal optimization problem for the illustrative example is defined as below:

\begin{equation}\label{eq:illustrative_nominal}
\begin{aligned}
\underset{x,y}{\text{max}} \; & - 2x^6 + 12.2x^5 - 21.2x^4 + 6.4x^3 + 4.7 x^2 - 12.74533x \\& - y^6 + 11y^5 - 43.3 y^4 + 74.8y^3 - 56.9y^2 + 11.43686y \\& + 4.1xy + 0.1x^2y^2 - 0.4xy^2 - 0.4x^2y + 12.66273\\
\text{s.t.} \;
&-1.0 \leq x \leq 3.5\\
&-0.5 \leq y \leq 4.5
\end{aligned}
\end{equation}
where, for the purposes of illustration, we assume the objective function and constraints represent some RTO objective such as profit and constraints such as quality specifications. Moreover, we assume the variables, $x, y \in \mathbb{R}$, are the states of the system to be controlled for which we wish to find set-points for the control layer by solving the RTO problem. However, as extensively discussed, the control layer must contend with implementation errors in the form of disturbances and noise. The goal is to find optimal set-points that exhibit robustness to these errors. To do this, we solve the equivalent constrained adversarially robust optimization problem, by re-writing the nominal problem in equation (\ref{eq:illustrative_nominal}) in the same form as the robust problem as per equation (\ref{eq:robust_constrained_prob}). Critically, for the adversarially robust problem, the uncertainty set, from which the possible implementation errors, $\boldsymbol{\Delta}\mathbf{x}$, are drawn is defined as:

\begin{equation}\label{eq:illustrative_uncertainty_set}
\mathcal{U} = \left\{\boldsymbol{\Delta}\mathbf{x} \;| 
 \; ||\boldsymbol{\Delta}\mathbf{x}||_2 \leq 0.3 \right\}
\end{equation}
where $\mathbf{x} = [x,y]^T \in \mathbb{R}^2$, are the states of the system as defined previously. Notice that $\Gamma$ is chosen to be $0.3$ in this case study, representing the maximum perturbation size to safeguard against. While chosen for the sake of exemplification, in reality, it is a crucial factor that requires careful consideration as it determines the desired level of robustness at the RTO layer. Specifically, this value should be chosen on the basis of the controller design and the robustness already available at the control layers. For instance, consider the scenario of temperature control in a vessel. The controller may already be designed to keep the temperature within $\pm 5^\circ$C of the set-point. In such a scenario, robustness beyond this range is not necessary since the controller already addresses it effectively. Consequently, these bounds from the control layer can be used to inform the value of $\Gamma$ at the RTO layer via the ARRTOC algorithm. By incorporating the controller design with RTO via ARRTOC in this way, we ensure that the set-points identified at the RTO layer strike a balanced level of conservativeness by avoiding excessive caution ($\Gamma$ too large) and undue leniency ($\Gamma$ too small). This characteristic stands as a significant advantage of the ARRTOC algorithm.

\begin{figure}
\begin{center}
\includegraphics[width=0.45\textwidth]{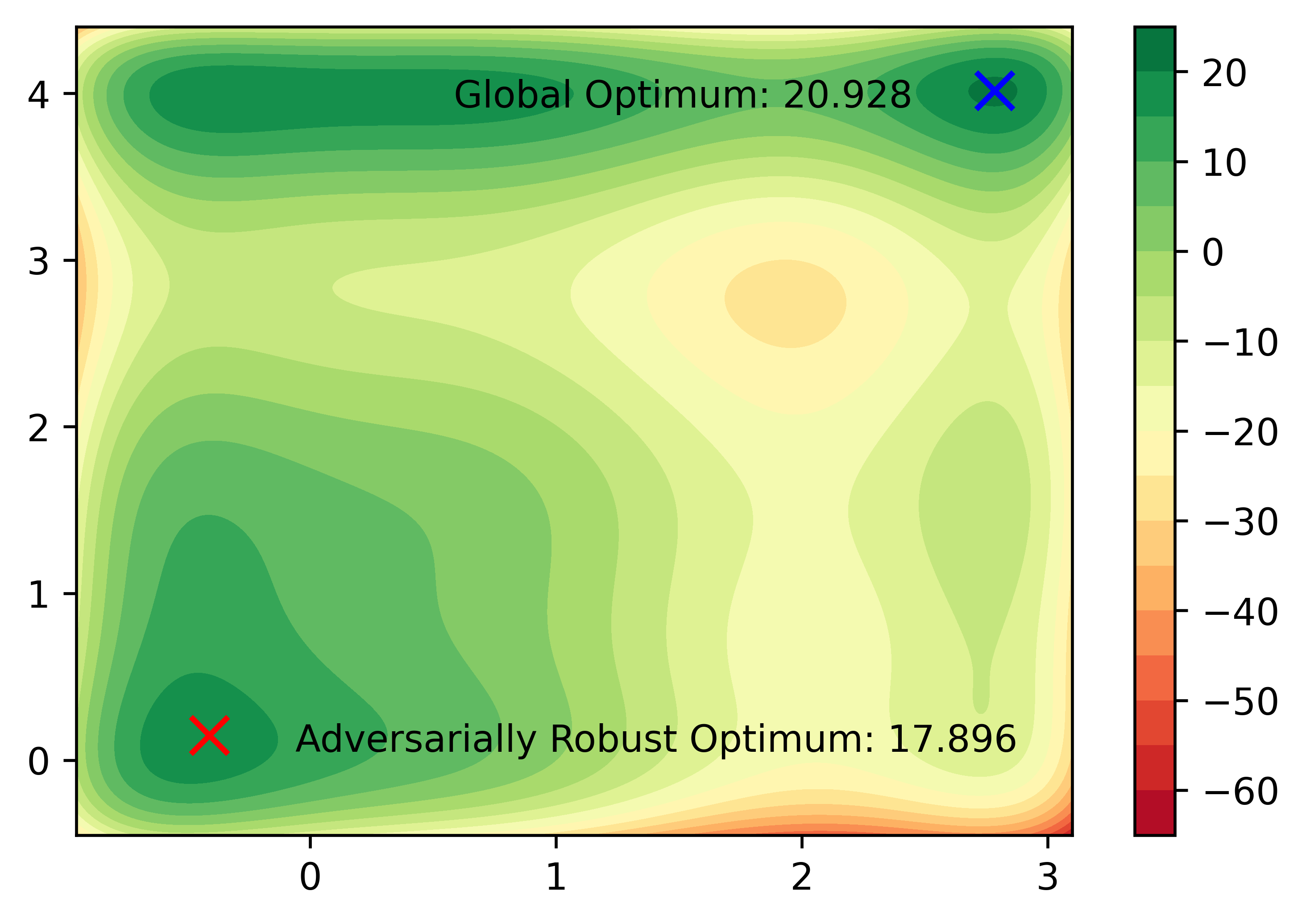} 
\caption{A contour plot of the nominal optimization problem from Eq. (\ref{eq:illustrative_nominal}). The global optimum is depicted as a blue cross while the adversarially robust optimum, given $\Gamma = 0.3$, is depicted as a red cross.} 
\label{fig:illustrative_objective_1}
\end{center}
\end{figure}

Continuing with the example, we solve the robust problem using the ARRTOC algorithm detailed in section \ref{sec:background_meth}. The results are depicted in Figure \ref{fig:illustrative_objective_1} which shows the objective function as a contour plot overlaying state-space. The global optimum set-point, $(2.78, 4.02)$, is denoted as a blue cross and the adversarially robust optimum set-point, $(-0.41, 0.15)$, is denoted as a red cross. The global optimum is found on a "narrow" peak of the objective function, making it sensitive to perturbations. In the context of RTO and control, this relates to the poor operability of this set-point \cite{Paulson2022, Gazzaneo2020, Georgiadis2002}. In contrast, the adversarially robust optimum is located on a "wide" peak of the objective function, suggesting that the objective value of this set point will be insensitive to perturbations when implemented at the control layers. The adversarially robust set-point demonstrates better operability properties.

\begin{figure}
     \centering
     \begin{subfigure}[b]{0.4\textwidth}
         \centering
         \includegraphics[width=\textwidth]{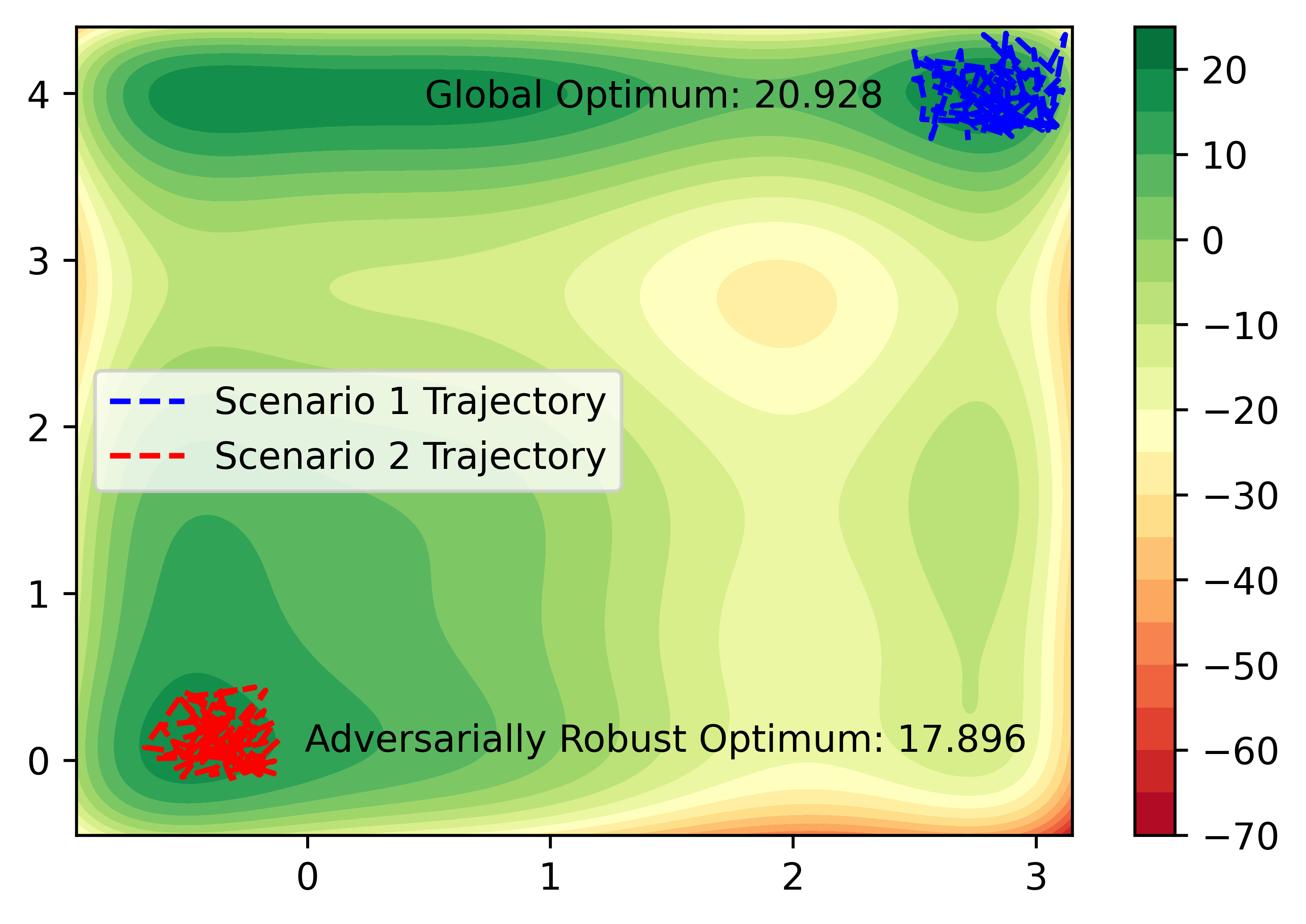}
         \caption{}
         \label{fig:illustrative_objective_2}
     \end{subfigure}
     \begin{subfigure}[b]{0.58\textwidth}
         \centering
         \includegraphics[width=\textwidth]{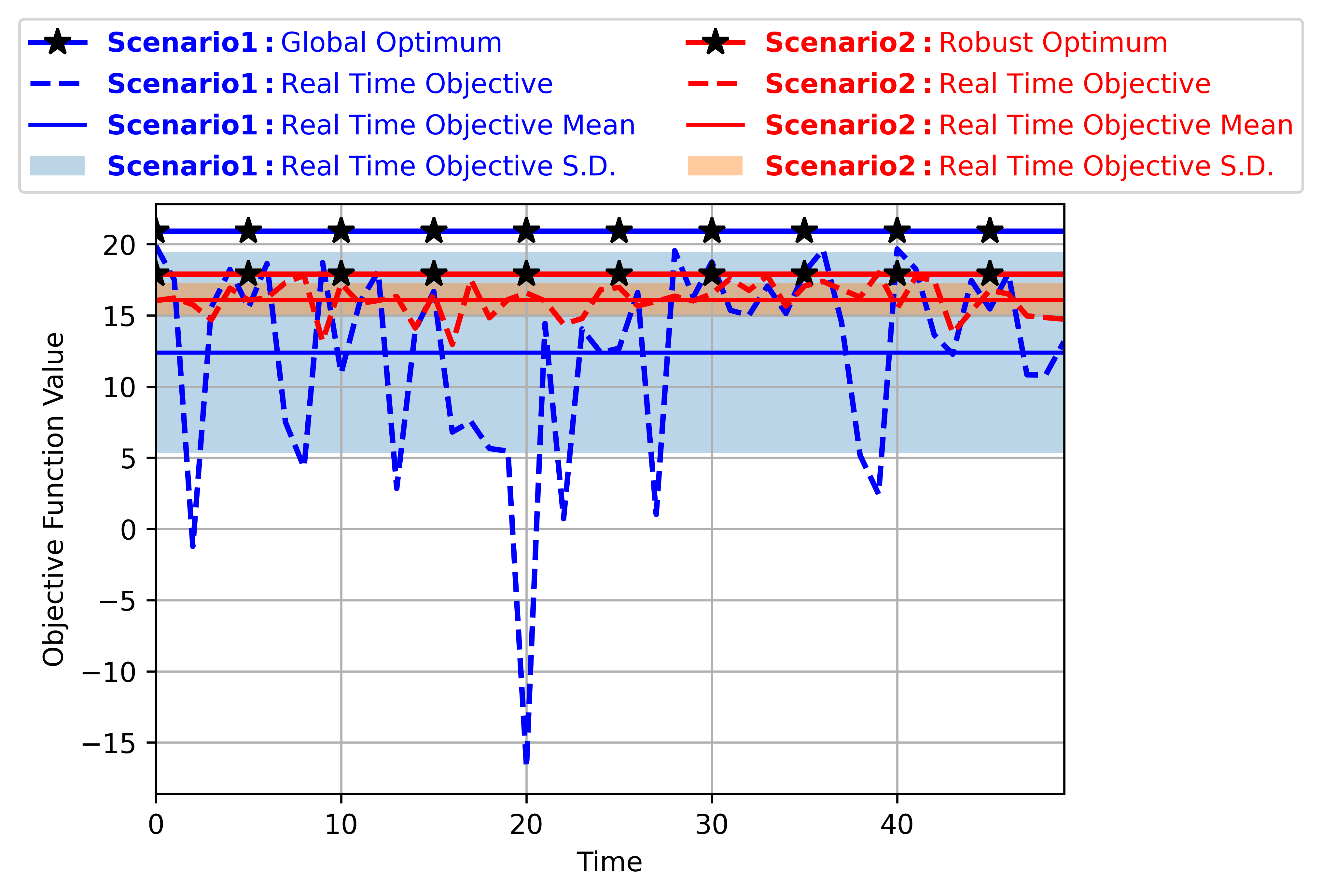}
         \caption{}
         \label{fig:illustrative_objective_3}
     \end{subfigure}
        \caption{(a): Two controller trajectories overlay the state-space objective function contour plot. Blue represents control around the global optimum set-point, while red indicates control around the adversarially robust set-point. (b): Dynamic performance of these trajectories is displayed in corresponding colours. This includes real-time objective value (coloured dashed line), average real-time objective (coloured solid line), real-time objective standard deviation (coloured hue), and expected optimum value (coloured solid line with black starred markers).}
        \label{fig:illustrative_control_performance}
\end{figure}

To underscore this particular point, we direct our attention to the hypothetical control scenarios illustrated in Figure \ref{fig:illustrative_control_performance}. We assume the controllers can maintain system control within $\pm 0.3$ units of the set-point, equivalent to the chosen $\Gamma$ value. In Figure \ref{fig:illustrative_objective_2}, we show the behaviour of the hypothetical controllers near the set-points in state-space for two scenarios. In scenario 1 (blue dashed line) the controller tracks the global optimum while in scenario 2 (red dashed line) the controller tracks the adversarially robust optimum. The dynamic performance of these two scenarios is shown in Figure \ref{fig:illustrative_objective_3}, which shows the real-time, average, and standard deviation of the objective value during control. In scenario 1, the real-time objective value (blue dashed line) fluctuates significantly due to system disturbances, occasionally dropping below zero. The mean objective value over time is $12.40$ (solid blue line), notably lower than the global optimum of $20.93$ (solid blue line with black starred markers), with a gap of over $40\%$. This suggests that the system struggles to operate at this point. The real-time objective's standard deviation (blue shaded region) stands at $7.05$. Clearly, this shows that, by design, the global optimum is not robust to adversarial perturbations. Conversely, in scenario 2, the real-time objective value (red dashed line) exhibits greater stability compared to scenario 1. The robust set-point ensures the mean objective value of $16.09$ (red solid line) is much closer to the adversarially robust optimum of $17.90$ (solid red line with black starred markers), with a gap of around $10\%$. The standard deviation (red shaded region) is notably smaller at $1.18$, contrasted with $7.05$ in scenario 1. In fact, due to the inherent robustness of this set-point, the mean objective value of scenario 2 exceeds scenario 1 by nearly $30\%$, highlighting the superior operability properties of the adversarially robust optimum. That is, paradoxically, operating at the adversarially robust optimum yields better average RTO objective performance than operating at the global optimum.

As previously noted, the critical parameter, which dictates the desired level of robustness required at the RTO level given the robustness already available at the control levels is the choice of $\Gamma$. In our current scenario, where the controllers were designed to maintain the system within $\pm 0.3$ units of the desired set-point, $\Gamma$ was set accordingly. Choosing a larger $\Gamma$ value would lead to excessively cautious set-points, assuming the controllers are less robust than designed. For instance, if the controllers were designed to keep the system within $\pm 0.1$ units of the desired set-point, we would not choose the adversarially robust set-point from Figure \ref{fig:illustrative_objective_1}, as it sacrifices optimality for excessive robustness. Instead, by re-solving the robust problem with $\Gamma = 0.1$, we find that the adversarially robust optimum and global optimum are equivalent. This highlights the importance of integrating RTO and control through the ARRTOC algorithm, allowing controller design to inform the required level of robustness at the RTO level. In some cases, no additional RTO-level robustness may be necessary as the nominal optimum is sufficiently robust given the controller design. In this illustrative example, we have explored hypothetical control scenarios that have simplified the selection of $\Gamma$. However, in practical situations, selecting $\Gamma$ can also be accomplished in a straightforward manner through empirical studies. This can be achieved through controller simulation studies or testing during deployment, both common industry practices \cite{Maddalena2020, Liporace2009}. These studies can be used to establish bounds on the level of robustness of the controllers to implementation errors which can subsequently be used to choose $\Gamma$. We employ this approach in our practical case studies in sections \ref{rd:bioreactor} and \ref{rd:evaporator}, using controller tuning simulations to guide $\Gamma$ selection.

\begin{figure}
\begin{center}
\includegraphics[width=0.7\textwidth]{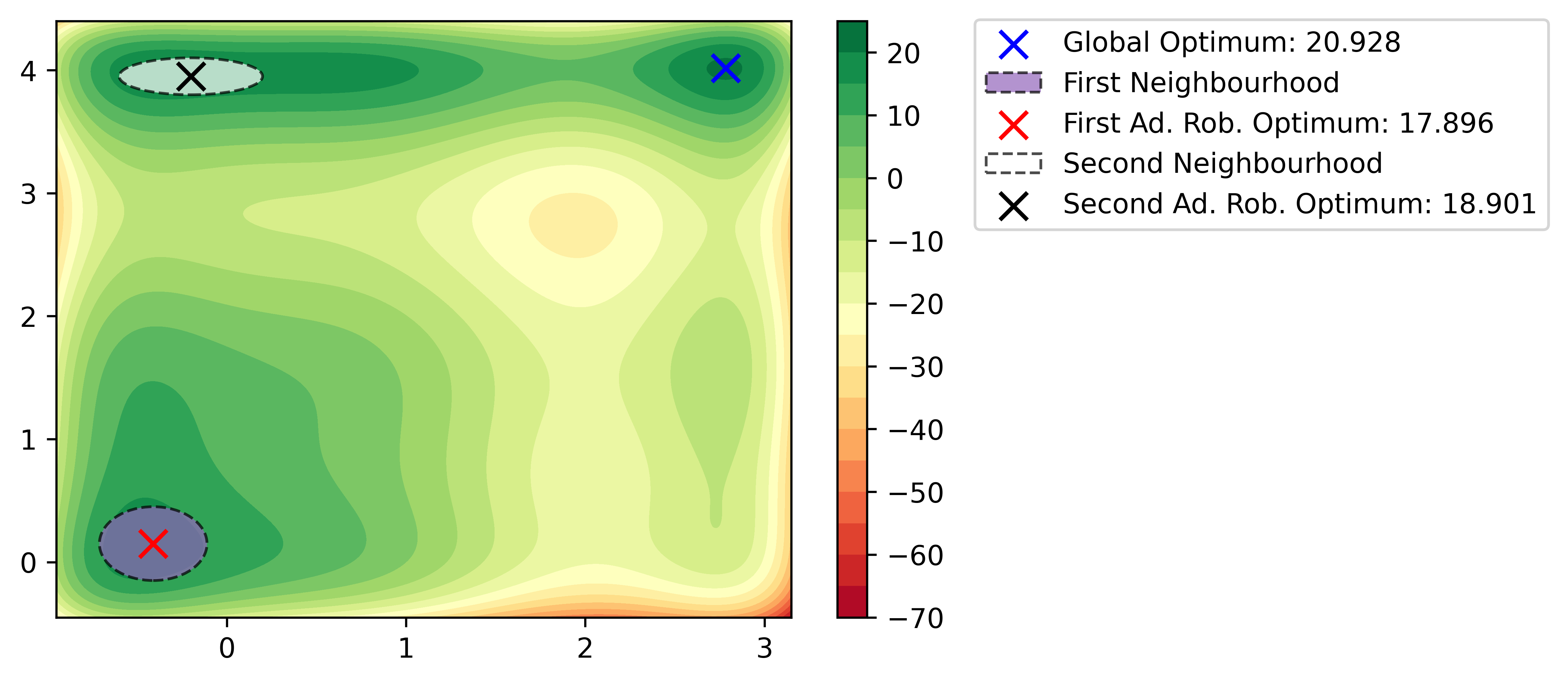} 
\caption{The objective function is displayed as a contour plot, with the global optimum (blue cross) and the first adversarially robust optimum (red cross) from Figure \ref{fig:illustrative_objective_1} depicted. The neighbourhood of the first adversarially robust optimum is depicted as a purple circle. Re-solving the robust problem with the uncertainty set in Eq. (\ref{eq:illustrative_general_uncertainty_set}) yields the second adversarially robust optimum (black cross), where the neighbourhood is shown as a white ellipse.} 
\label{fig:illustrative_objective_5}
\end{center}
\end{figure}

Finally, we use the illustrative example to highlight an additional insight relevant to the practical case studies of sections \ref{rd:bioreactor} and \ref{rd:evaporator}. Recall in section \ref{constrained:prob_def}, we introduced a generalisation of the uncertainty set, equation (\ref{eq:ellipsis_uncertainty_set}), more suited to RTO and control problems. We argued that this generalisation could be used to find adversarially robust set-points which are tailored to the level of robustness required for each of the states by accounting for different maximum perturbations, $\Gamma_i$, for each state independently. To highlight the ease of integrating the new uncertainty set with the ARRTOC algorithm and to illustrate the point visually, we re-solve the adversarially robust version of the illustrative example with the general uncertainty set defined as:

\begin{equation}\label{eq:illustrative_general_uncertainty_set}
\mathcal{U} = \left\{\boldsymbol{\Delta}\mathbf{x} \; \middle | 
 \; \frac{{\Delta x}^2}{{\Gamma_x}^2} + \frac{{\Delta y}^2}{{\Gamma_y}^2} \leq 1 \right\}
\end{equation}
where, as before, $\boldsymbol{\Delta}\mathbf{x} = [\Delta x, \Delta y]^T \in \mathbb{R}^2$, are the possible implementation errors of the system. Most importantly, we define $\Gamma_x = 0.4$ and $\Gamma_y = 0.15$. This encodes the idea that we wish to find set-points which are more robust to perturbations in state $x$ than state $y$. The results of solving this new robust problem are depicted in Figure \ref{fig:illustrative_objective_5}. The blue cross represents the global optimum of the nominal problem. The red cross represents the first adversarially robust optimum from Figure \ref{fig:illustrative_objective_1} found using the uncertainty set defined in equation (\ref{eq:illustrative_uncertainty_set}) with the neighbourhood of the point represented as a purple hue. The new adversarially robust optimum found using the general definition of the uncertainty set as per equation (\ref{eq:illustrative_general_uncertainty_set}), is represented as a black cross. Its neighbourhood is represented as a white hue which in this case takes the shape of an ellipsis with semi-axis lengths of $\Gamma_x = 0.4$ and $\Gamma_y = 0.15$. Notice that the new adversarially robust optimum set-point is tailored to the specific maximum level of perturbations expected for each of the states individually. This insight finds practical application in section \ref{rd:evaporator}.

\subsection{Continuous Bioreactor Process}\label{rd:bioreactor}

Continuous automated bioprocess operation is crucial for maximizing the efficiency and productivity of bioprocesses, which are characterised by limited profitability \cite{Ho2019, Oliveira2018, Zydney2015, Chen2018, Mowbray2021, Krausch2022}. Continuous bioreactors are the primary revenue generators in these processes, utilising microorganisms to produce high-value products. However, their operational sensitivity poses a fundamental challenge, with perhaps the biggest issue being wash-out, an abrupt operational failure which disrupts the performance of the bioreactor \cite{Oliveira2018, Wang1999, Doran2012, Banu2021}. 

\begin{figure}
\begin{center}
\includegraphics[width=0.5\textwidth]{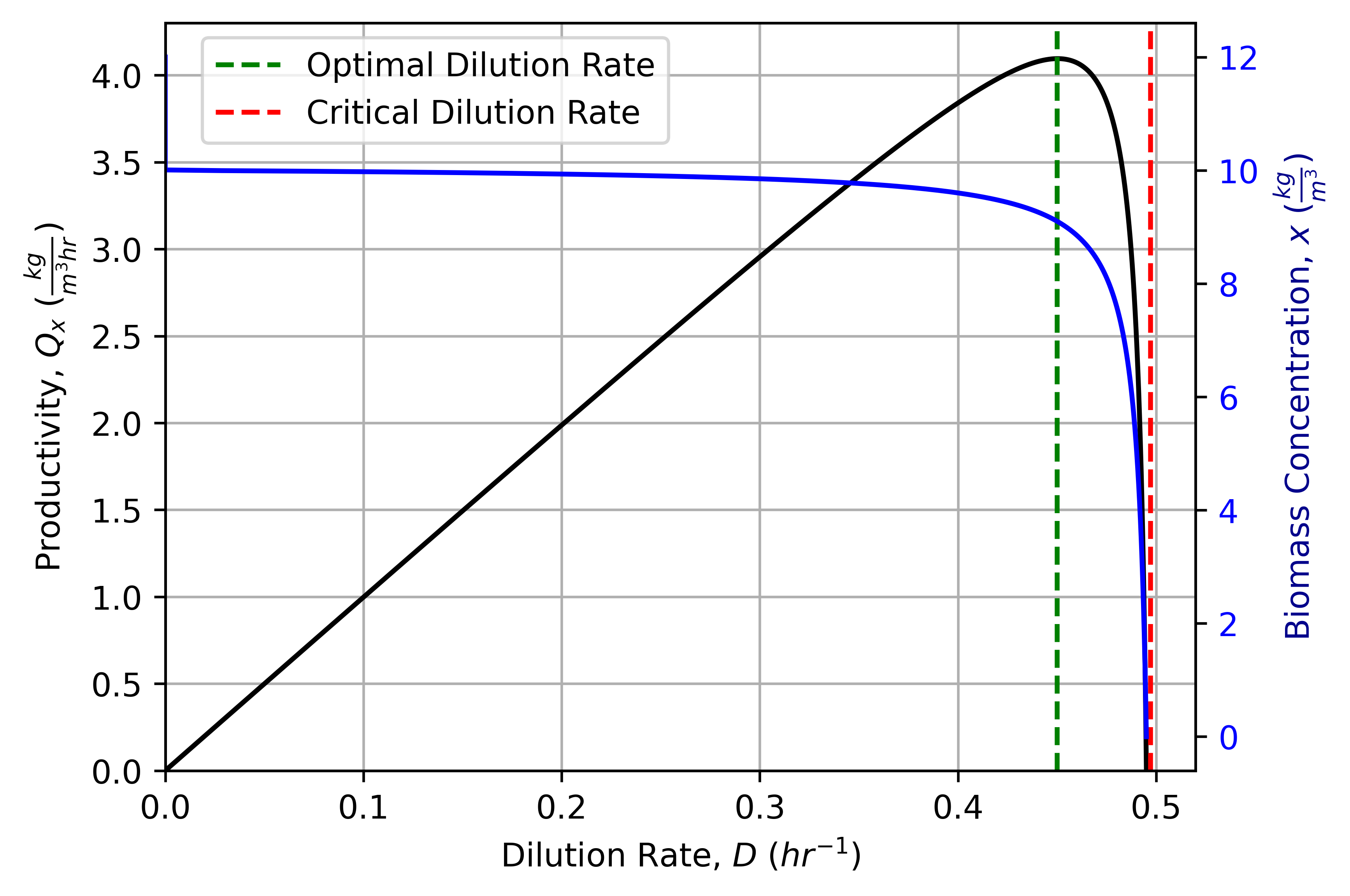} 
\caption{Plot of the productivity vs dilution rate (black curve) and the biomass concentration vs dilutation rate (blue curve). The green dashed line indicates the optimal dilution rate in terms of productivity. The red dashed line indicates the critical dilution rate which represents the point at which wash-out occurs.} 
\label{fig:bioreactor_curves}
\end{center}
\end{figure}

Wash-out occurs when high flow rates in a bioreactor detach microorganisms from the reactor walls due to strong shear forces, resulting in an abrupt and unrecoverable loss of biomass. This leads to a complete halt in biochemical reactions, causing the bioreactor's efficiency and productivity to plummet to zero. Typically, this catastrophic event necessitates a complete shutdown of the operation, introducing new biomass, and initiating a lengthy bioreactor start-up process, which can last from days to weeks to allow for cell acclimation \cite{Zydney2015, Chen2018, Ajbar2001, Doran2012}. This contradicts the core goal of continuous bioprocess operation, which aims to avoid lengthy start-up processes associated with batch or fed-batch operations \cite{Zydney2015, Chen2018}. However, to achieve optimal bioreactor productivity, high throughput conditions are essential; otherwise, the bioreactor's productivity remains too low for profitability. Studies on continuous bioreactor operation have shown that profitability is typically achieved when operating at flow rates or dilution rates close to their optimal values for productivity \cite{Oliveira2018, Zydney2015, Chen2018, Wang1999, Ajbar2001}. The operational challenge lies in finding a delicate balance between optimality and operability. Figure \ref{fig:bioreactor_curves} illustrates this challenge, showcasing the productivity (black curve) and biomass concentration (blue curve) with respect to the dilution rate. The optimal dilution rate (green dashed vertical line) sits dangerously close to the critical dilution rate (red dashed vertical line), where wash-out occurs and productivity sharply drops to zero. Any disturbances, noise, or estimation errors at the control layers that push the dilution rate beyond the optimal value can result in partial or total wash-out. Consequently, striking the right balance between optimality and operability is crucial for maximizing reactor productivity and profitability while minimizing the risk of catastrophic events like wash-out.

This challenge poses a significant bottleneck to achieving continuous bioprocess operation. Researchers aim to address this issue through various heuristic solutions, including alterations to bioreactor design \cite{Loosdrecht1995, Escudie2011}, or optimization of media and nutrients to improve adhesion \cite{Marbelia2014, Zaiat1996}. However, these changes often result in additional costs or adverse effects on productivity. Alternatively, attempts to enhance control algorithms for added robustness tend to complicate the already intricate control strategies for bioreactors, rendering them impractical \cite{Wang1999, Escudie2011, Banu2021, Zhai2020}. Consequently, industrial practitioners often resort to rule-of-thumb approaches (such as a naive back-off from the optimum). These approaches have to be employed on an ad-hoc basis due to significant variations in the depicted curves (Figure \ref{fig:bioreactor_curves}) across systems, often more complex than the ideal conditions assumed \cite{Wang1999, Banu2021, Zhai2020, Doran2012}. In contrast, the ARRTOC algorithm precisely addresses processes requiring a finely balanced trade-off between optimality and operability, without introducing additional complexity to the control layers. In this section, we illustrate this using a simplified, ideal bioreactor case study. It's worth noting that while our example is straightforward, ARRTOC's principles can be adapted to more complex bioreactors, provided an accurate steady-state model of the system is available. To demonstrate this adaptability, we consider a scenario where limited knowledge of the bioreactor necessitates the use of a data-driven Gaussian Process (GP) model, constructed from steady-state data to approximate the biomass productivity which serves as the objective function in this case study. Moreover, the use of a GP model as the objective function serves as an initial step to demonstrate that the ARRTOC algorithm can accommodate models of various types whether data-driven or even simulation-based. This highlights the algorithm's readiness for model adaptation methodologies, such as adaptive gaussian process learning or model-parameter adaptation, such that the model can be updated and adapted online \cite{Ahmed2021, Zagorowska2023, Chanona2021, Chachuat2009, Mercangoz2008}.

For the case study, we consider an ideal continuous bioreactor with one inlet stream and one outlet stream. The bioreactor consists of two main components, a biomass ($x$) and substrate ($s$) concentration (units of $\frac{\text{kg}}{\text{m}^3}$). The dynamics of the two species are defined below \cite{Doran2012}:

\begin{equation}\label{eq:bioreactor_dynamics}
\begin{aligned}
\frac{\mathrm{d}x}{\mathrm{d}t} &= (\mu - D)x \\
\frac{\mathrm{d}s}{\mathrm{d}t} &= D(s_i - s) - \frac{\mu x}{Y_{xs}} \\
\mu &= \frac{\mu_{\text{max}}s}{K_s + s}\\
\end{aligned}
\end{equation}
where $x$ and $s$ are the state variables, $\mathbf{x} = [x, s]^\intercal \in \mathbb{R}^2$, with the biomass concentration being the controlled variable, while the manipulated input variable, $u = D$, is the dilution rate (units of $\text{hr}^{-1}$). We assume that the disturbance variable is the inlet concentration of substrate, $s_i$, with a mean inlet concentration of $20\frac{\text{kg}}{\text{m}^3}$ and a standard deviation of $2\frac{\text{kg}}{\text{m}^3}$ i.e. $s_i \sim N(20, 2)$. All other variables are treated as fixed parameters with their values given in Table \ref{param_table_1}.

\begin{table}
\caption{Table of parameter values used for equation (\ref{eq:bioreactor_dynamics}).}
\begin{center}
\begin{tabular}{ c  c  c  c }
\hline
Parameter & Description & Value & Units \\ \hline
$\mu_{\text{max}}$ & Maximum specific growth rate & 0.5 & 1/hr \\ 
$Y_{xs}$ & Biomass yield from substrate & 0.5 & kg/kg \\ 
$K_s$ & Substrate constant & 0.2 & kg/m$^3$ \\ \hline
\end{tabular}
\label{param_table_1}
\end{center}
\end{table}

In this case study, we assume the true plant dynamics are described by equation (\ref{eq:bioreactor_dynamics}), of which we have no knowledge, which is typically the case for most bioprocesses. To solve the RTO problem, an approximation of the objective function is required. In this case study, we employ a Gaussian Process (GP) model to approximate the objective function, biomass productivity, using steady-state data. The selection of a GP model is motivated by its suitability for data-driven modeling in situations characterised by limited data, which is commonly encountered in bioprocess systems \cite{Ahmed2021, Zagorowska2023, Chanona2021, Rasmussen2005}. The objective function, which we wish to maximize, of the bioreactor is the volumetric biomass productivity, $Q_x$, (units of $\frac{\text{kg}}{\text{m}^3\text{hr}}$) which is defined as:

\begin{equation}\label{eq:bioreactor_productivity}
Q_x = Dx
\end{equation}
As mentioned, the biomass productivity of the reactor, subject to steady-state conditions, defines the objective function for the RTO problem where the decision variable is the biomass concentration set-point. The nominal problem is given as:

\begin{equation}\label{eq:bioreactor_nominal}
\begin{aligned}
\underset{x}{\text{max}} \; & Q_x\\
\text{s.t.} \;
& \text{Equation (\ref{eq:bioreactor_dynamics}) at steady-state}
\end{aligned}
\end{equation}
However, as noted, we assume no knowledge of the steady-state plant. Instead, we assume that steady-state measurements of the dilution rate and biomass concentration exist which can then be used to calculate the biomass productivity of the reactor using equation (\ref{eq:bioreactor_productivity}). In this case study, we perform a steady-state system identification experiment by taking 20 measurements of the simulated system at steady-state at equally spaced values of the dilution rate in the range between $0\text{hr}^{-1}$ and $0.5\text{hr}^{-1}$. Once the system reaches steady-state, the corresponding biomass concentrations are measured which can then be used to calculate the biomass productivity.

\begin{figure}
\begin{center}
\includegraphics[width=0.7\textwidth]{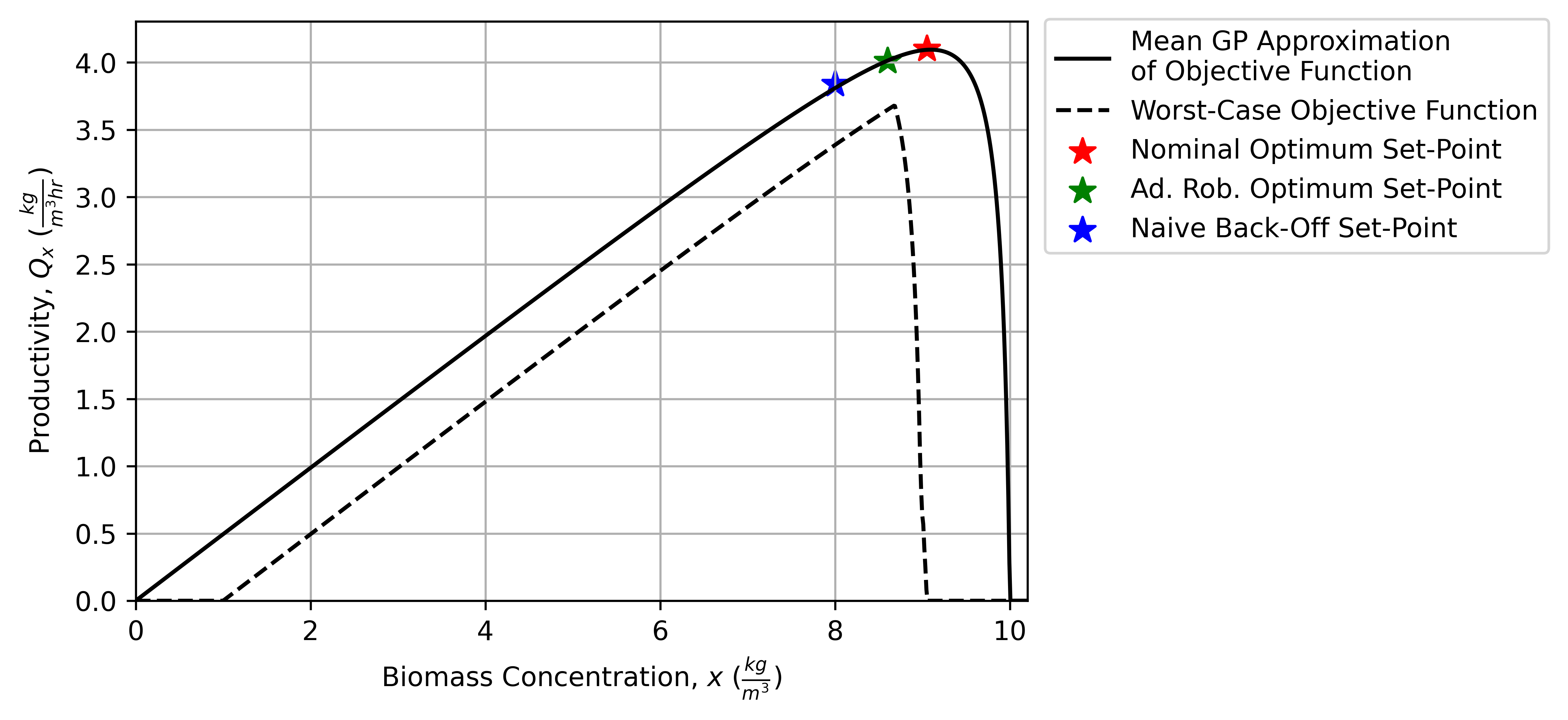} 
\caption{The GP approximation of the objective function is represented by a solid black curve. The nominal optimum set-point, maximizing the black curve, is marked with a red star. A blue star denotes the naive back-off set-point, calculated by subtracting the expected maximum disturbance level from the nominal optimum. The adversarially robust set-point is shown as a green star, achieved with $\Gamma = 1.0$. The dashed black curve illustrates the worst-case objective function, with the adversarially robust optimum corresponding to its maximizer.} 
\label{fig:bioreactor_three_optima}
\end{center}
\end{figure}

Given the collected biomass concentration and productivity data, which we denote as $\mathbf{x}^k \in \mathbb{R}^k$ and $\mathbf{Q}_\mathbf{x}^k \in \mathbb{R}^k$, respectively, where $k$ is the number of samples taken ($k = 20$ in our case), a data-driven approximation of the objective function, $Q_{x,\text{GP}}(x) : \mathbb{R} \to \mathbb{R}$, can be obtained. This approximation represents the biomass productivity as a function of the biomass concentration, $x$, under steady-state conditions, and it is achieved by fitting a GP on the input, $\mathbf{x}^k$, and output datasets, $\mathbf{Q}_\mathbf{x}^k$. The GP can be thought of as a multivariate Gaussian distribution over functions from which a function, $g : \mathbb{R} \to \mathbb{R}$, can be sampled:

\begin{equation}\label{eq:gp_bioreactor}
\begin{aligned}
&g(x) \sim \mathcal{GP}(m(\cdot), k(\cdot, \cdot))\\
&Q_{x,\text{GP}}(x) = m(\cdot)
\end{aligned}
\end{equation}
where $m(\cdot)$ and $k(\cdot, \cdot)$, represent the mean function and covariance function of the GP trained with the input-output dataset \cite{Rasmussen2005}. The GP regression problem was solved using GPy \cite{Erickson2018}. The corresponding approximation of the  objective function is depicted as a black solid curve in Figure \ref{fig:bioreactor_three_optima}. As previously noted, the RTO goal is to find the biomass concentration set-point which gives the optimal productivity. Solving the nominal optimization problem, gives an optimal set-point of approximately $9.0\frac{\text{kg}}{\text{m}^3}$ which gives a productivity of $4.1\frac{\text{kg}}{\text{m}^3\text{hr}}$. This is depicted as a red star in Figure \ref{fig:bioreactor_three_optima}. However, as extensively discussed, this choice is likely to be a poor one when implemented at the control layers due to implementation errors, in this case in the form of disturbances to the inlet substrate concentration. Of course, given this, our goal is to solve the RTO problem in the adversarially robust optimization problem setting. However, before doing this, we need a good choice of the critical parameter $\Gamma$ which dictates the maximum size of the perturbations we wish to protect against. To do this we must first consider the underlying controller design to inform the level of robustness required at the RTO layer. 

For this case study, we employ a tuned PI controller to regulate the biomass concentration, $x$, at the desired set-point by manipulating the dilution rate, $D$. Disturbances enter the system in the form of inlet substrate concentration variations, with a mean value of $20\frac{\text{kg}}{\text{m}^3}$ and a standard deviation of $2\frac{\text{kg}}{\text{m}^3}$: $s_i \sim N(20, 2)$. We tuned the PI controllers through trial and error, ultimately settling on $KC = 0.1$ for the proportional gain and $KI = 0.05$ for the integral gain. These settings maintained the biomass concentration within approximately $\pm 1.0 \frac{\text{kg}}{\text{m}^3}$ of the set-point during various simulations. Consequently, we use $\Gamma = 1.0$ to solve the equivalent adversarially robust optimization problem, yielding an optimum biomass concentration of approximately $8.5\frac{\text{kg}}{\text{m}^3}$ at a productivity of $4.0\frac{\text{kg}}{\text{m}^3\text{hr}}$. In Figure \ref{fig:bioreactor_three_optima}, the dashed black curve represents the worst-case objective function for $\Gamma = 1.0$, with the adversarially robust optimum (maximizer of this curve) indicated by a green star.

Finally, in Figure \ref{fig:bioreactor_three_optima}, apart from the nominal optimum (red star) and the adversarially robust optimum (green star) we also include, for comparison purposes, a commonly employed rule-of-thumb, a naive back off of $1.0 \frac{\text{kg}}{\text{m}^3}$ (based on the expected level of disturbances at the control layer) from the nominal optimum, as a blue star. Insights from Figure \ref{fig:bioreactor_three_optima} reveal valuable information. The nominal optimum, while offering maximum productivity, may be operationally risky, as slight disturbances could lead to wash-out. In fact, the worst-case objective function value at the nominal optimum drops to zero, suggesting that a perturbation of just $1.0 \frac{\text{kg}}{\text{m}^3}$ could cause wash-out. Conversely, the rule-of-thumb approach, valued for its robustness, sacrifices optimality significantly. This is a known limitation in the bioprocess literature, where such approaches struggle to account for system-specific features, including asymmetry and multi-modality in the productivity-concentration curves \cite{Wang1999, Ajbar2001, Banu2021}. This makes rule-of-thumb approaches inadequate as they fail to capture these properties of real systems. In contrast, the adversarially robust optimum achieves a balance between optimality and operability by leveraging knowledge of the objective function's specific features (in this ideal case, the asymmetric nature of the objective function about the nominal optimum). With an accurate model, the ARRTOC algorithm can effectively harness system-specific characteristics, making it a promising approach for real bioreactors which show more complex behaviours.

\begin{figure}
     \centering
     \begin{subfigure}[b]{0.32\textwidth}
         \centering
         \includegraphics[width=\textwidth]{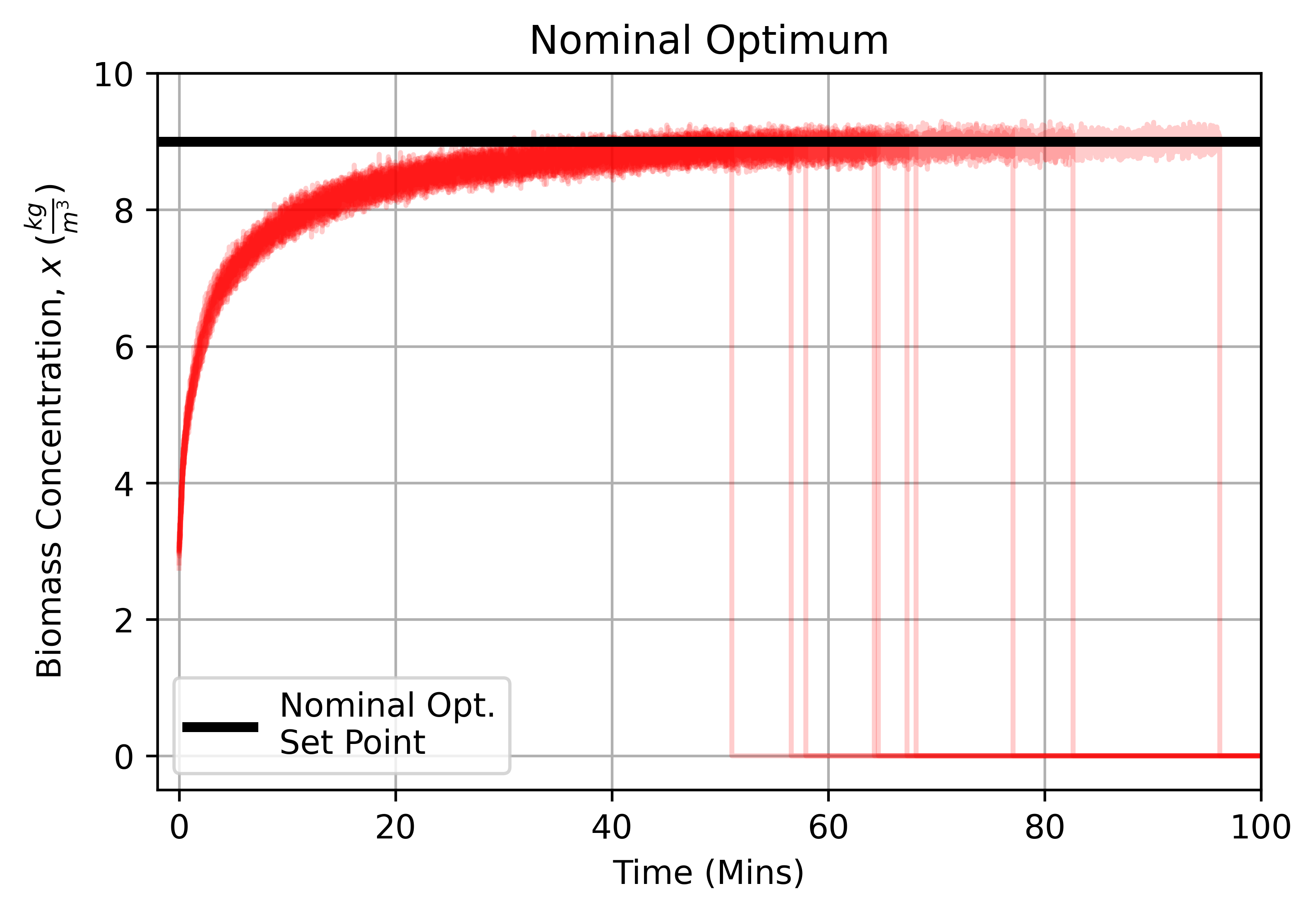}
         \caption{}
         \label{fig:bioreactor_control_performance_global}
     \end{subfigure}
     \begin{subfigure}[b]{0.32\textwidth}
         \centering
         \includegraphics[width=\textwidth]{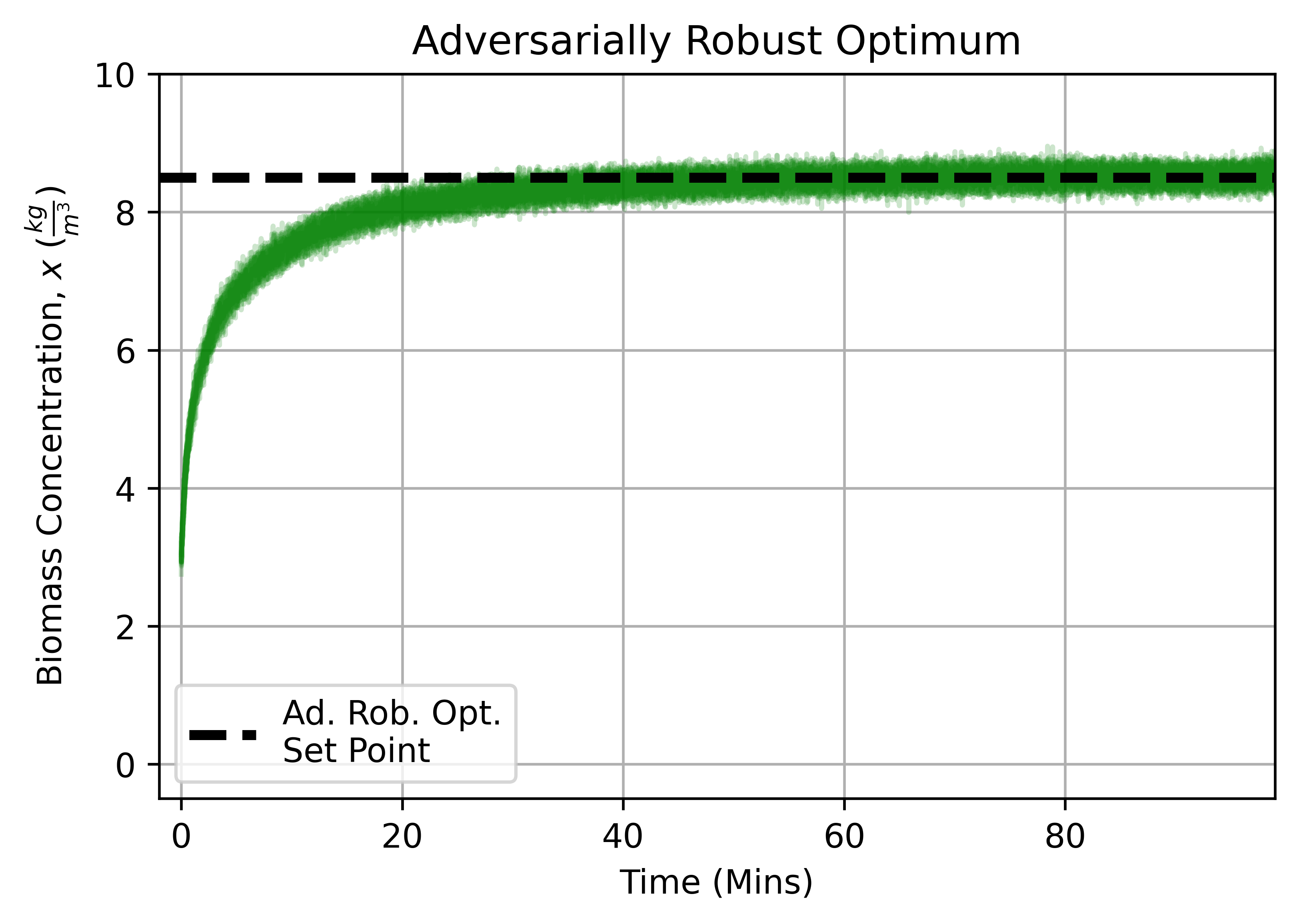}
         \caption{}
         \label{fig:bioreactor_control_performance_aro}
     \end{subfigure}
     \begin{subfigure}[b]{0.32\textwidth}
         \centering
         \includegraphics[width=\textwidth]{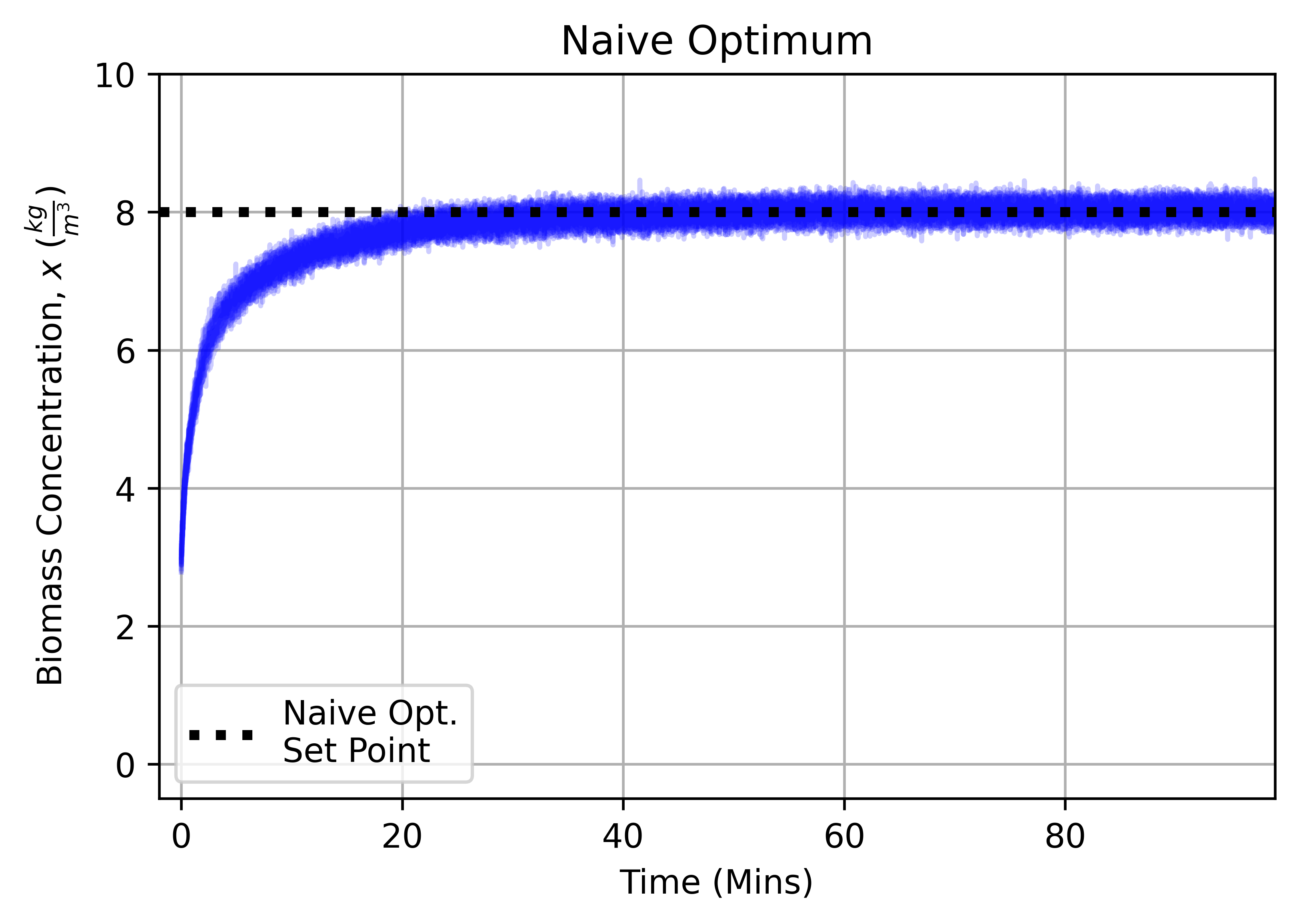}
         \caption{}
         \label{fig:bioreactor_control_performance_naive}
     \end{subfigure}
        \caption{Each case presents 10 simulation runs. (a): Controller performance at the nominal optimum set-point, where all runs experience wash-out. (b): Controller performance at the adversarially robust set-point, where no wash-out occurs in any simulation. (c): Controller performance at the naive back-off set-point, again no wash-out is observed.}
        \label{fig:bioreactor_control_performance}
\end{figure}

\begin{figure}
\begin{center}
\includegraphics[width=0.5\textwidth]{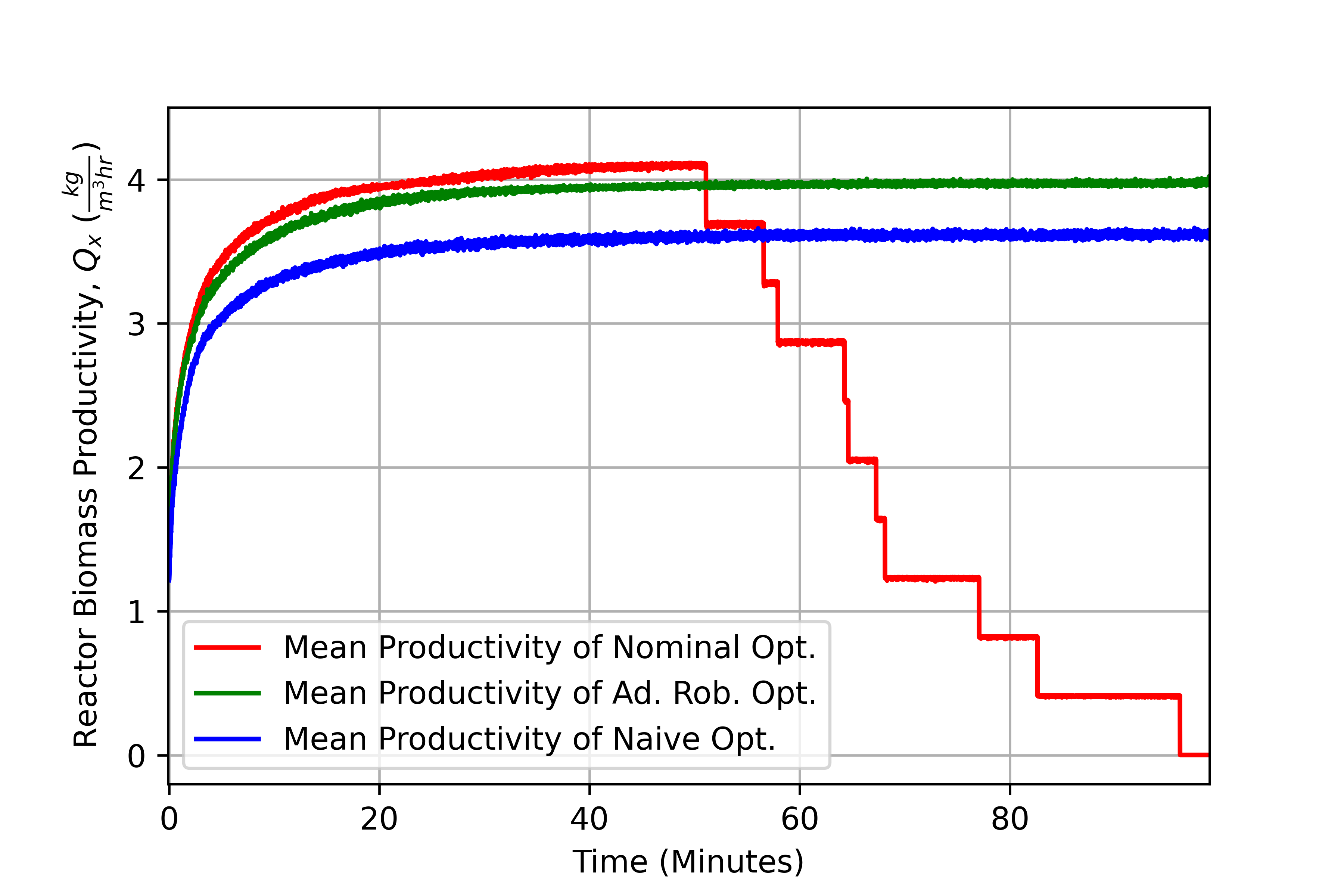} 
\caption{The mean real-time productivity is depicted for each scenario with the same colour scheme as Figure \ref{fig:bioreactor_control_performance}. The mean productivity at the nominal optimum declines to zero due to wash-out in all simulation runs. While both are robust to wash-out, the mean productivity at the adversarially robust set-point exceeds that of the naive set-point.} 
\label{fig:bioreactor_prod_performance}
\end{center}
\end{figure}

To empirically test our hypotheses, we conducted three sets of controller experiments for each set-point, aiming to control the biomass concentration from an initial value of $3.0\frac{\text{kg}}{\text{m}^3}$ to the desired set-point using the previously defined PI controller. Each of the experiments involve 10 simulation runs. The controller performance for the three sets of experiments is depicted in Figure \ref{fig:bioreactor_control_performance}. For the nominal optimum set-point of $9.0\frac{\text{kg}}{\text{m}^3}$, all simulation runs led to wash-out, exposing its inoperability (Figure \ref{fig:bioreactor_control_performance_global}). In contrast, the adversarially robust optimum set-point of $8.5\frac{\text{kg}}{\text{m}^3}$ and the naive back-off set-point of $8.0\frac{\text{kg}}{\text{m}^3}$ proved to be robust, with no instances of wash-out observed (Figures \ref{fig:bioreactor_control_performance_aro} and \ref{fig:bioreactor_control_performance_naive}). However, the performance contrast between the naive and adversarially robust optima is evident in Figure \ref{fig:bioreactor_prod_performance}, displaying averaged real-time biomass productivity across the 10 simulation runs for each experiment. Firstly, note, as expected, the mean real-time productivity for the nominal optimum is initially the largest, but quickly decays to zero due to wash-out in the various runs. More interestingly, while both the adversarially robust and naive back-off set-points proved to be robust to wash-out, here we can see the difference between the two solutions. The naive back-off set-point sacrifices too much productivity for robustness while, in comparison, the adversarially robust set-point yields $10\%$ higher productivity. In the context of narrow profit margins in bioprocess economics, this difference is significant \cite{Oliveira2018, Chen2018, Banu2021}. 

\subsection{Evaporator Process}\label{rd:evaporator}

In this section, we present a case study focussing on the application of ARRTOC to a multi-loop control strategy for an evaporator process \cite{Seborg2011, Kam2000, Yadav2010}. The choice of the evaporator process is driven by its widespread use in industries such as chemicals, pharmaceuticals, and food processing, where precise control is crucial to meet product quality standards \cite{Glover2004, Shallan2019, Diaz2023}. However, evaporators are known for their sensitivity to operational perturbations, stemming from both their design and the inherent complexities and nonlinear dynamics of the process \cite{Glover2004, Diaz2023}. Furthermore, the multi-loop structure yields unique dynamic effects when subjected to disturbances and noise, highlighting the need for enhanced robustness. Our case study explores the dynamic interactions among the control loops, illustrating how ARRTOC leverages them for superior control performance. Importantly, this does not require additional control-layer complexity or modifications, rendering it a practical solution for industrial practitioners.

We examine an evaporator process with a feed stream containing two components: a non-volatile solute dissolved in a volatile solvent. Heat is supplied via a steam line to evaporate the solvent. The evaporator has both vapour and liquid outlets. Our control strategy employs three PI controllers. The primary controlled variable is the solute composition ($x_B$) in the liquid product stream which is controlled by adjusting the steam temperature ($T_S$). This control loop exhibits sluggish response and non-minimum phase characteristics, making the product composition disturbance-sensitive \cite{Seborg2011, Kam2000, Yadav2010}. In addition to solute composition, we also control two other important variables: liquid level ($h$) for safety reasons, regulated by adjusting the liquid product stream flow rate ($B$), and pressure ($P$), which also has a major influence on the vapour-liquid equilibrium of the evaporator, controlled by the vapour flow rate ($D$). Both control loops are very fast acting. The system dynamics are given by the following set of differential-algebraic system of equations:

\begin{equation}\label{eq:evaporator_dynamics}
\begin{aligned}
\frac{\mathrm{d}h}{\mathrm{d}t} &= \frac{1}{A_T c} (F - B - D) \\
\frac{\mathrm{d}x_B}{\mathrm{d}t} &= \frac{1}{A_T h c} (Fx_F - Bx_B) - \frac{x_B}{h}\frac{\mathrm{d}h}{\mathrm{d}t}\\
\frac{\mathrm{d}\rho}{\mathrm{d}t} &= \frac{M_W}{V_{\text{vap}}} (E - D) - \frac{\rho}{V_{\text{vap}}} \frac{\mathrm{d}V_{\text{vap}}}{\mathrm{d}t}\\
P &= \frac{\rho R T}{M_W}\\
E &= \frac{UA_S}{\Delta H_v} (T_S - T)\\
V_{T} &= V_{\text{vap}} + A_T h\\
T &= \frac{B_{\text{ant}}}{A_{\text{ant}} - \log\left(\frac{P}{133.322}\right)} - C_{\text{ant}} + 273.15\\
\end{aligned}
\end{equation}
where $h$ is the level of the liquid in the evaporator (units of m), $x_B$ is the mole fraction of the solute in the product stream, and $P$ is the pressure of the evaporator (units of Pa). These three variables represent the state variables, $\mathbf{x} = [x_B, h, P]^\intercal \in \mathbb{R}^3$. Note that the vapour mass density, $\rho$ (units of $\frac{\text{kg}}{\text{m}^3}$) is an explicit state variable in the system of equations. However, as pressure is one of the controlled variables, it can be derived from vapor mass density using the ideal gas law as in equation (\ref{eq:evaporator_dynamics}). The same principle applies to other equation of state variables affecting vapour-liquid equilibrium: the evaporator vapour volume, $V_{\text{vap}}$ (units of $\text{m}^3$) and the evaporator temperature $T$ (units of K). The controlled variables are all of the state variables, $\mathbf{x}$, which are controlled by the manipulated variables of the the steam temperature, $T_S$ (units of K), the product flow rate, $B$ (units of $\frac{\text{mol}}{\text{s}}$) and the vapour flow rate, $D$ (units of $\frac{\text{mol}}{\text{s}}$) i.e. $\mathbf{u} = [T_S, B, D]^\intercal \in \mathbb{R}^3$. We assume that the disturbance variables are the feed flow rate $F$ (units of $\frac{\text{mol}}{\text{s}}$) and the feed composition, $x_F$. The exact characteristics of these disturbance variables will be discussed during the simulation of the dynamics.  Lastly, $E$ denotes the evaporation rate ($\frac{\text{mol}}{\text{s}}$), derived from an energy balance assuming negligible sensible heat effects compared to latent heat. All other variables are considered fixed parameters, as listed in Table \ref{param_table_2}.

\begin{table}
\caption{Table of parameter values used for equation (\ref{eq:evaporator_dynamics}).}
\begin{center}
\begin{tabular}{ c  c  c  c }
\hline
Parameter & Description & Value & Units \\ \hline
$A_T$ & Cross-sectional area of evaporator tank & 100 & m$^2$ \\ 
$c$ & Concentration of solute in solvent & 10 & mol/m$^3$ \\ 
$M_W$ & Molecular weight of solvent & 0.078 & kg/mol \\
$R$ & Universal gas constant & 8.3145 & J/molK \\
$U$ & Overall heat transfer coefficient & 1000 & W/m$^2$K \\
$A_S$ & Steam heat transfer area & 50 & m$^2$ \\
$\Delta H_v$ & Enthalpy of vaporisation of solvent & 30800 & J/mol \\
$V_T$ & Volume of evaporator tank & 1000 & m$^3$ \\
$A_{\text{ant}}$ & Antoine Coefficient 1 & 6.87987 & - \\
$B_{\text{ant}}$ & Antoine Coefficient 2 & 1196.76 & - \\
$C_{\text{ant}}$ & Antoine Coefficient 3 & 219.161 & - \\\hline
\end{tabular}
\label{param_table_2}
\end{center}
\end{table}

The RTO goal is to find the set-points for the states, $\mathbf{x}$, which maximize the profit of the process subject to operational and safety constraints. The RTO profit objective is given below:

\begin{equation}\label{eq:evaporator_objective}
J_P = C_\text{product}Bx_B - C_\text{feed}F - C_\text{energy}T^{1.5} - C_\text{tank}h^{2}
\end{equation}
where $J_P$ is the profit of the process (units of $\frac{\text{\$}}{\text{s}}$) which consists of the product revenue, $C_\text{product}Bx_B$, minus the cost of the feed stream, $ C_\text{feed}F$, the cost of energy, $C_\text{energy}T^{1.5}$ and the cost of maintaining a liquid capacity in the tank, $C_\text{tank}h^{2}$. The prices are provided in Table \ref{param_table_3}.

\begin{table}
\caption{Table of cost values used for equation (\ref{eq:evaporator_objective}).}
\begin{center}
\begin{tabular}{ c  c  c  c }
\hline
Parameter & Description & Value& Units \\ \hline
$C_\text{product}$ & Price of product stream & $11.875x_B - 1.875$ & \$/mol\\ 
$C_\text{feed}$ & Cost of feed stream & 0.04 & \$/mol\\ 
$C_\text{energy}$ & Proxy cost of energy based on temperature & 0.01 & \$/K$^{1.5}$ \\
$C_\text{tank}$ & Proxy cost of maintaining liquid capacity based on liquid level & 0.75 & \$/m$^{2}$ \\\hline
\end{tabular}
\label{param_table_3}
\end{center}
\end{table}

The safety and operational constraints of the process necessitate that the steam temperature can safely range between $400$ and $450$K. The liquid level must be between $2$ and $8$m for safety purposes. Evaporator pressure should stay within $0.05$-$0.5$MPa, aligned with the evaporator design limits. Finally, the product stream mole fraction should not exceed a specification upper bound of $0.9$. 

\begin{figure}
     \centering
     \begin{subfigure}[b]{0.45\textwidth}
         \centering
         \includegraphics[width=\textwidth]{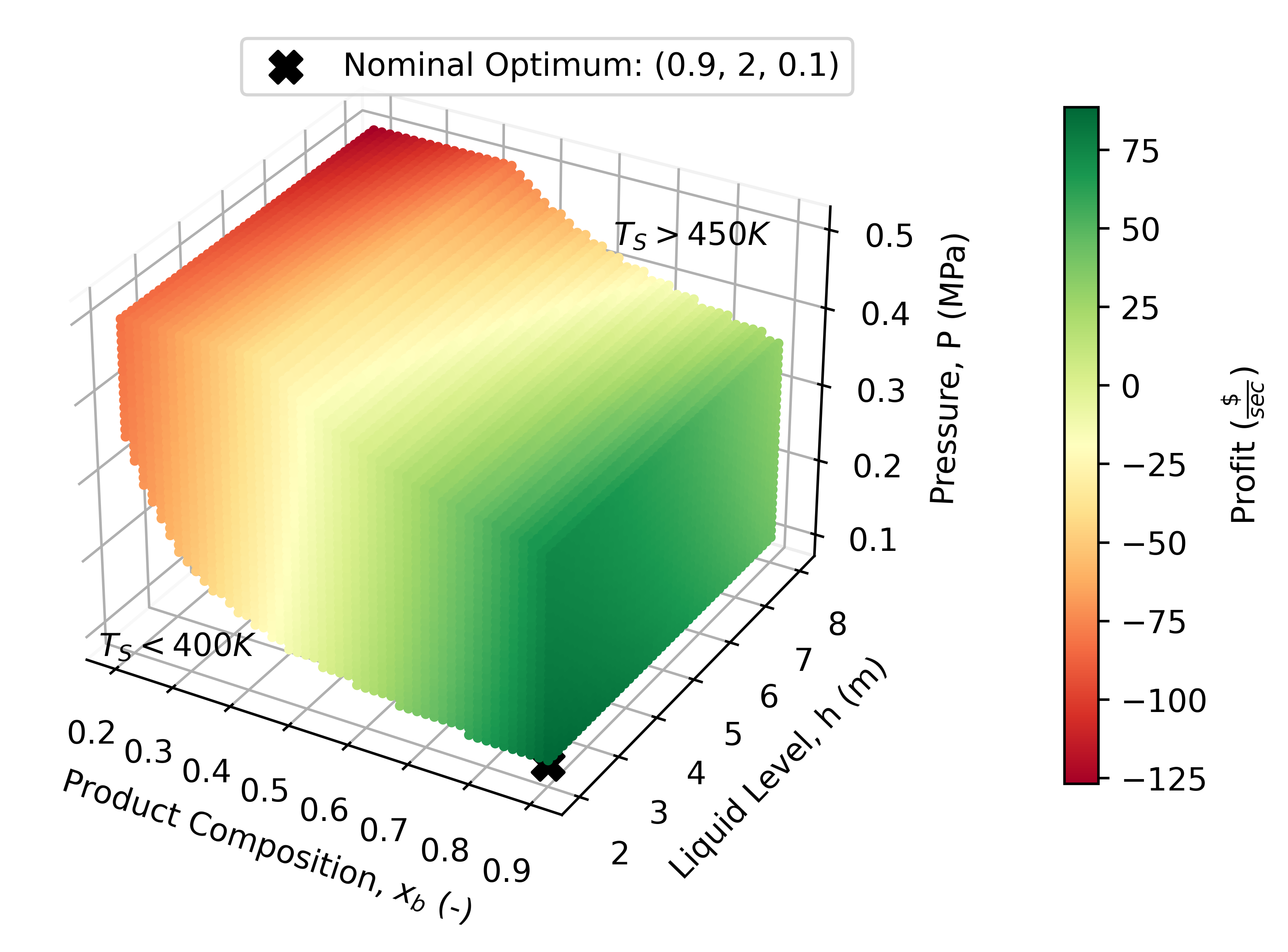}
         \caption{}
         \label{fig:evaporator_objective_view_1}
     \end{subfigure}
     \begin{subfigure}[b]{0.45\textwidth}
         \centering
         \includegraphics[width=\textwidth]{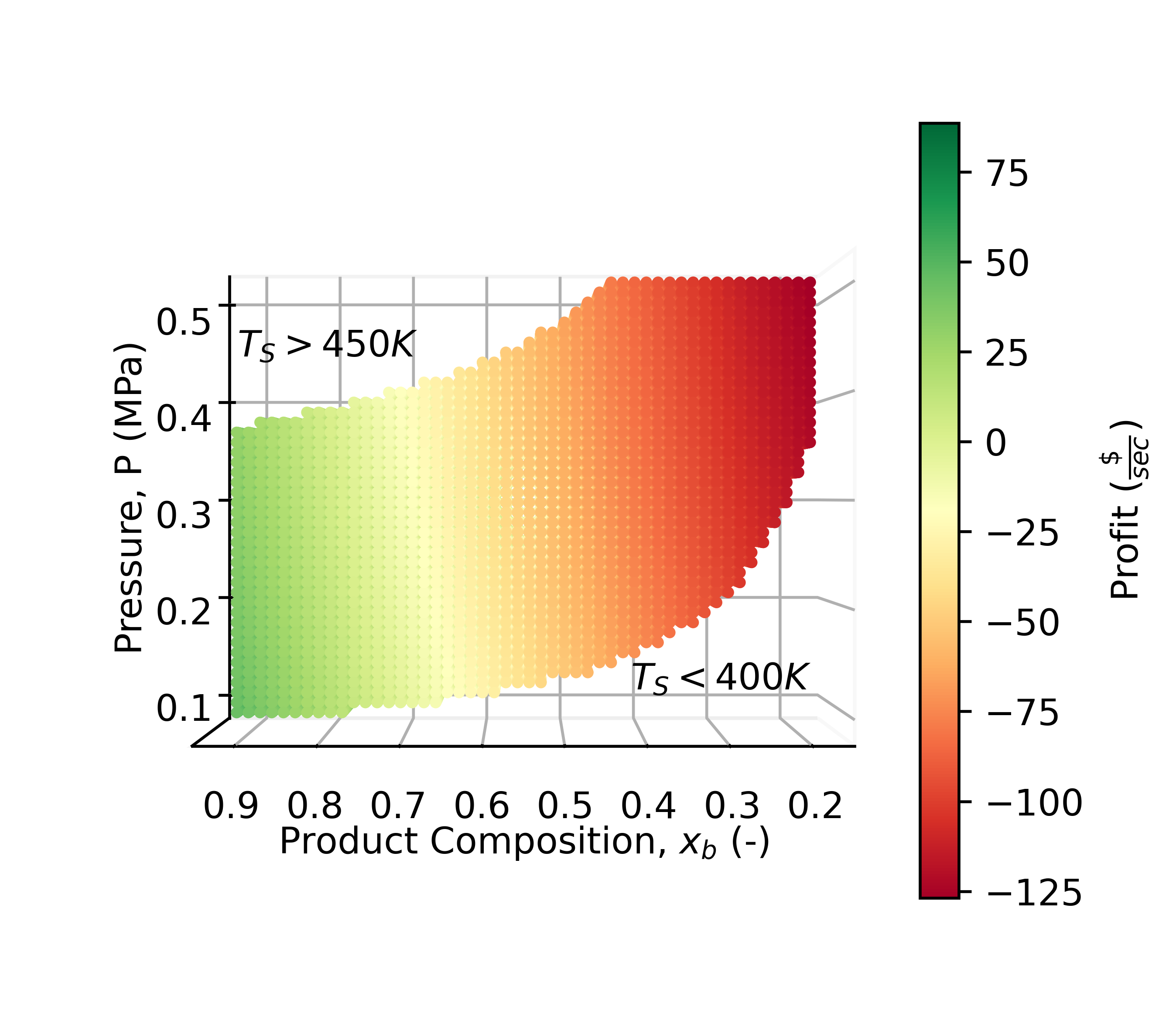}
         \caption{}
         \label{fig:evaporator_objective_view_2}
     \end{subfigure}
        \caption{Different views of the objective function and constraints of the nominal optimization problem from Eq. (\ref{eq:evaporator_nominal}). The objective function (profit) is represented as a colour gradient from red to green indicating low and high profit respectively. In (a) the nominal optimum set-point of $(x_B^* = 0.9, h^* = 2\text{m}, P^* = 0.1\text{MPa})$ is depicted as a black cross.}
        \label{fig:evaporator_3d_objective}
\end{figure}

Given the objective and the constraints, the nominal RTO problem is defined as follows:

\begin{equation}\label{eq:evaporator_nominal}
\begin{aligned}
\underset{x_B, h, P}{\text{max}} \; & J_P\\
\text{s.t.} \;
& \text{Equation (\ref{eq:evaporator_dynamics}) at steady-state}\\
& 400\text{K} \leq T_S \leq 450\text{K} \\
& 2\text{m}  \leq h \leq 8\text{m} \\
& 0.05\text{MPa}  \leq P \leq 0.5\text{MPa} \\
& x_B \leq 0.9
\end{aligned}
\end{equation}
Figure \ref{fig:evaporator_3d_objective} visually presents the objective function and constraints. The three axes represent the decision variables, and the colour gradient (red to yellow to green) indicates the objective value (profit) from low to high. Constraint violations are shown in colourless areas, signifying regions where the steam temperature falls below $400$K or exceeds $450$K, as depicted in Figure \ref{fig:evaporator_3d_objective}.

Solving the nominal RTO problem yields an optimum of $\frac{\text{\$}}{\text{s}}89.03$ for a set-point of $(x_B^* = 0.9, h^* = 2\text{m}, P^* = 0.1\text{MPa})$, represented by a black cross in Figure \ref{fig:evaporator_objective_view_1}. Notably, this optimum coincides with the intersection of three active constraints. However, maintaining consistent constraint satisfaction in the presence of disturbances and noise at the control layers presents a significant challenge, prompting our proposal of using ARRTOC. 

\begin{figure}
     \centering
     \begin{subfigure}[b]{0.4\textwidth}
         \centering
         \includegraphics[width=\textwidth]{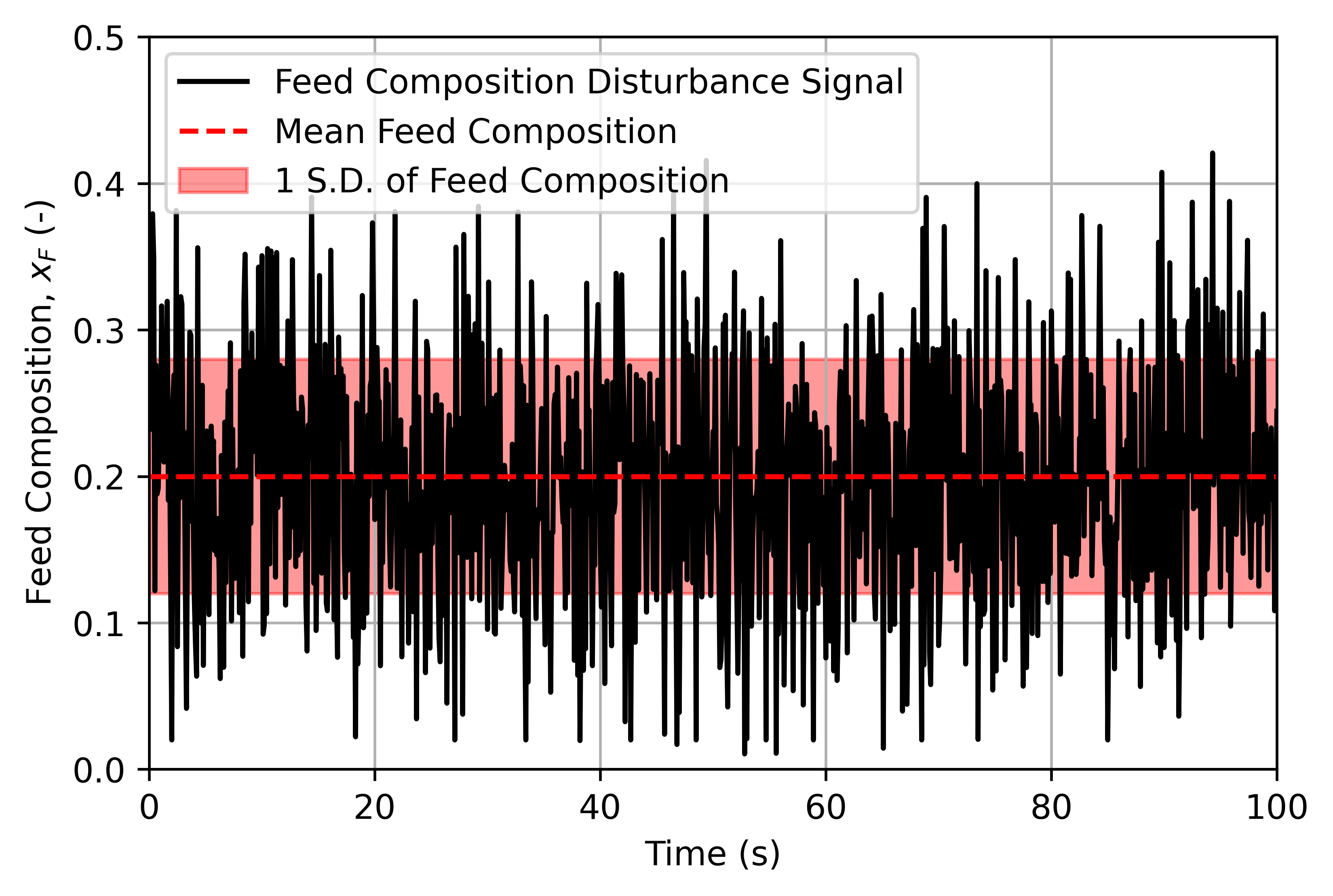}
         \caption{}
         \label{fig:composition_disturbance}
     \end{subfigure}
     \begin{subfigure}[b]{0.4\textwidth}
         \centering
         \includegraphics[width=\textwidth]{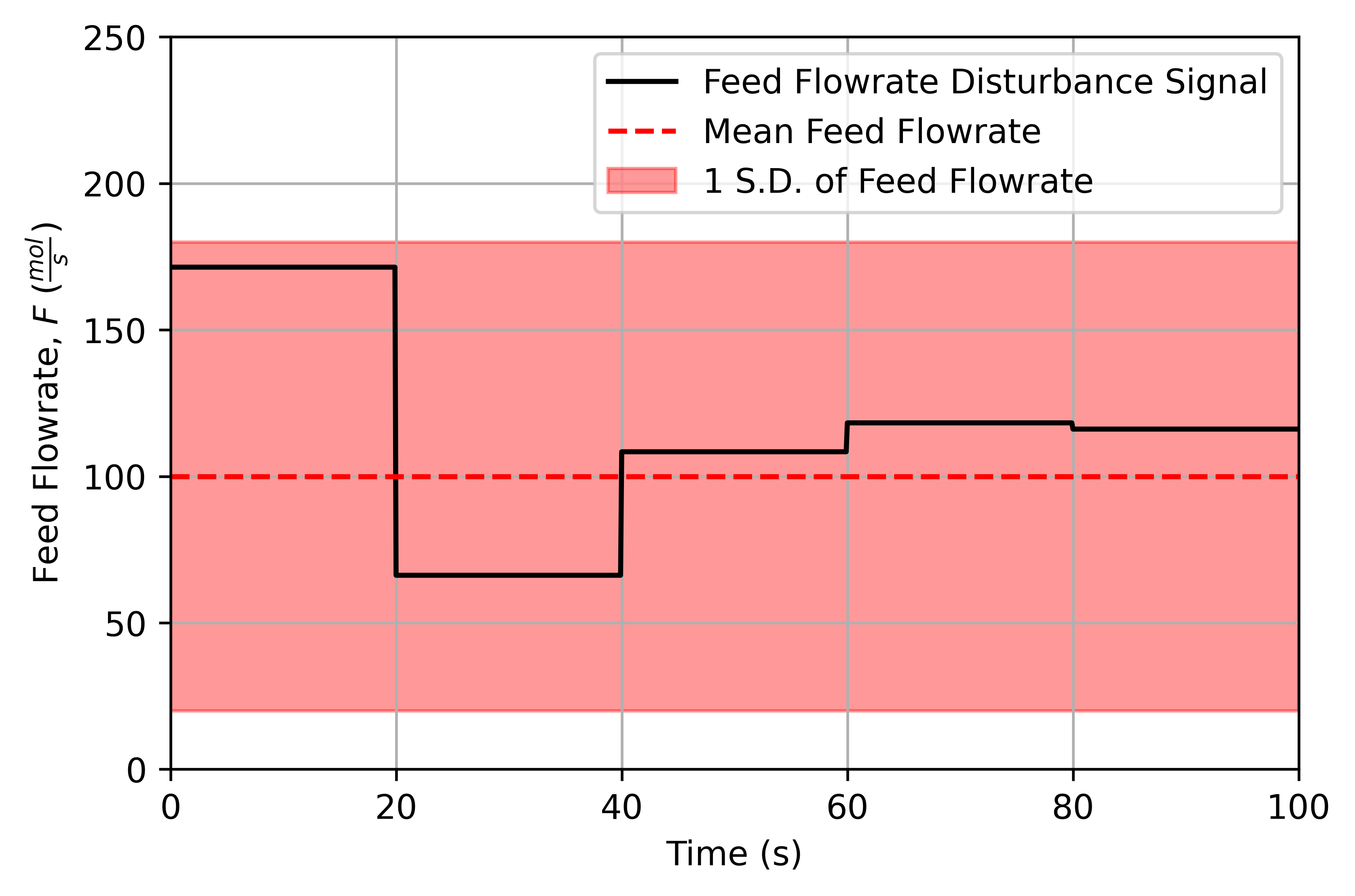}
         \caption{}
         \label{fig:feed_disturbance}
     \end{subfigure}
        \caption{(a): Feed composition disturbance signal (black curve). The mean (0.2) is indicated by a red dashed horizontal line, and the standard deviation (0.08) is represented by a red hue. (b): Feed flowrate disturbance signal (black curve). The mean (100 $\frac{\text{mol}}{\text{s}}$) is denoted with a red dashed horizontal line, and the standard deviation (80 $\frac{\text{mol}}{\text{s}}$) is denoted by a red hue.}
        \label{fig:disturbance_signals}
\end{figure}

To demonstrate the advantages of ARRTOC, we evaluate seven different controller designs with varying degrees of robustness to disturbances. We achieve this by tuning the underlying PI controllers to produce seven different controller settings each with different robustness characteristics. These controllers are tuned while the system experiences disturbances in both the feed flow rate and composition. We tune the controllers around an arbitrarily chosen set-point of $(x_B = 0.7, h = 5\text{m}, P = 0.1\text{MPa})$. This set-point is chosen to avoid biasing the results by tuning around any of the optimal set-points. In our controller tuning simulations, we assume the following disturbance characteristics: the feed composition follows a normal distribution i.e. $x_F \sim N(0.2, 0.08)$, with high frequency, shown in Figure \ref{fig:composition_disturbance}. The feed flowrate disturbance is a less frequent step disturbance, with the step values also sampled from a normal distribution i.e. $F \sim N(100, 80)$. We assume 5 step changes over the length of the simulation. This disturbance signal is depicted in Figure \ref{fig:feed_disturbance}.

The multi-loop controllers were tuned using a modified version of the classic sequential relay auto-tuning method combined with a derivative-free optimizer (Py-BOBYQA) to minimize the summed normalised integral of time-weighted absolute error (ITAE) for each loop \cite{Shen1994, Hang2002, Marchetti2002, Cartis2019, Cartis2022}. This was performed as follows: Initially, all loops except the fastest $P-D$ control loop were set to manual mode. Using the BOBYQA optimizer with a budget of $N$ iterations (which varies for each controller setting), we determine $KC_{PD}$ and $KI_{PD}$ for this loop, minimizing its ITAE. Then, the next fastest $h-B$ loop is switched to automatic mode. We jointly optimize $KC_{PD}$, $KI_{PD}$, $KC_{hB}$, and $KI_{hB}$ using the BOBYQA optimizer (with the same budget of $N$), using the sum of the normalised ITAE from both loops as the performance metric. Importantly, we initialise the $P-D$ loop gains as the optimal values from the previous iteration. The same process is repeated for the final $x_B-T_S$ loop, optimizing all six gains together with the sum of the normalised ITAE for all three loops used as the performance metric to be minimized. Again, the initial values of the gains for the $P-D$ and $h-B$ loops were taken to be the optimal values from the previous iteration. 

The above was repeated for 7 distinct controller designs, each with varying BOBYQA solver budgets, $N$, ranging from $10, 25, 50, 100, 500, 750$ and $1000$ iterations. This allows us to consider 7 different settings with differing degrees of robustness to disturbances as measured by their respective summed normalised ITAE metric. In theory, larger budgets enhance robustness by optimizing controller gains through more iterations. However, it is important to acknowledge that in practice, this may not hold true for two primary reasons. Firstly, the controller tunings were conducted around a specifically chosen set point of $(x_B = 0.7, h = 5\text{m}, P = 0.1\text{MPa})$, and although the selected gains may be "best" for this particular set point, they may not be as effective when applied to other operating conditions as we plan to do. Lastly, the summed ITAE for each of the loops under disturbances scenarios is a valid but superficial measure of robustness to disturbances. Nevertheless, we consider these limitations reasonable and reflective of real-world complexities. In fact, such uncertainties can be viewed as implementation errors stemming from the imperfect realisation process of the controller, which can also be accounted for by ARRTOC via the parameter $\Gamma$.

\begin{figure}
     \centering
     \begin{subfigure}[b]{0.32\textwidth}
         \centering
         \includegraphics[width=\textwidth]{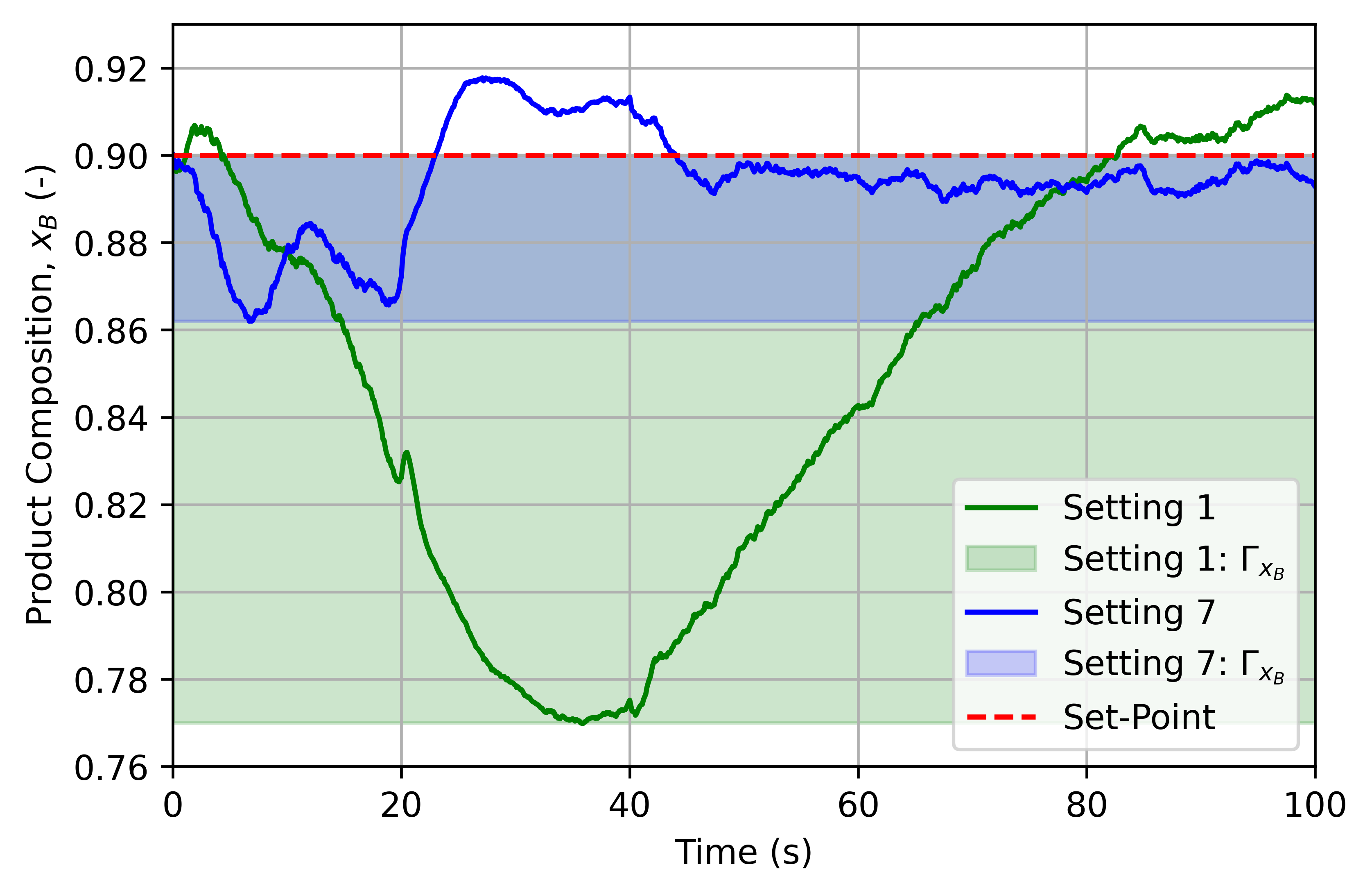}
         \caption{}
         \label{fig:composition_control}
     \end{subfigure}
     \begin{subfigure}[b]{0.32\textwidth}
         \centering
         \includegraphics[width=\textwidth]{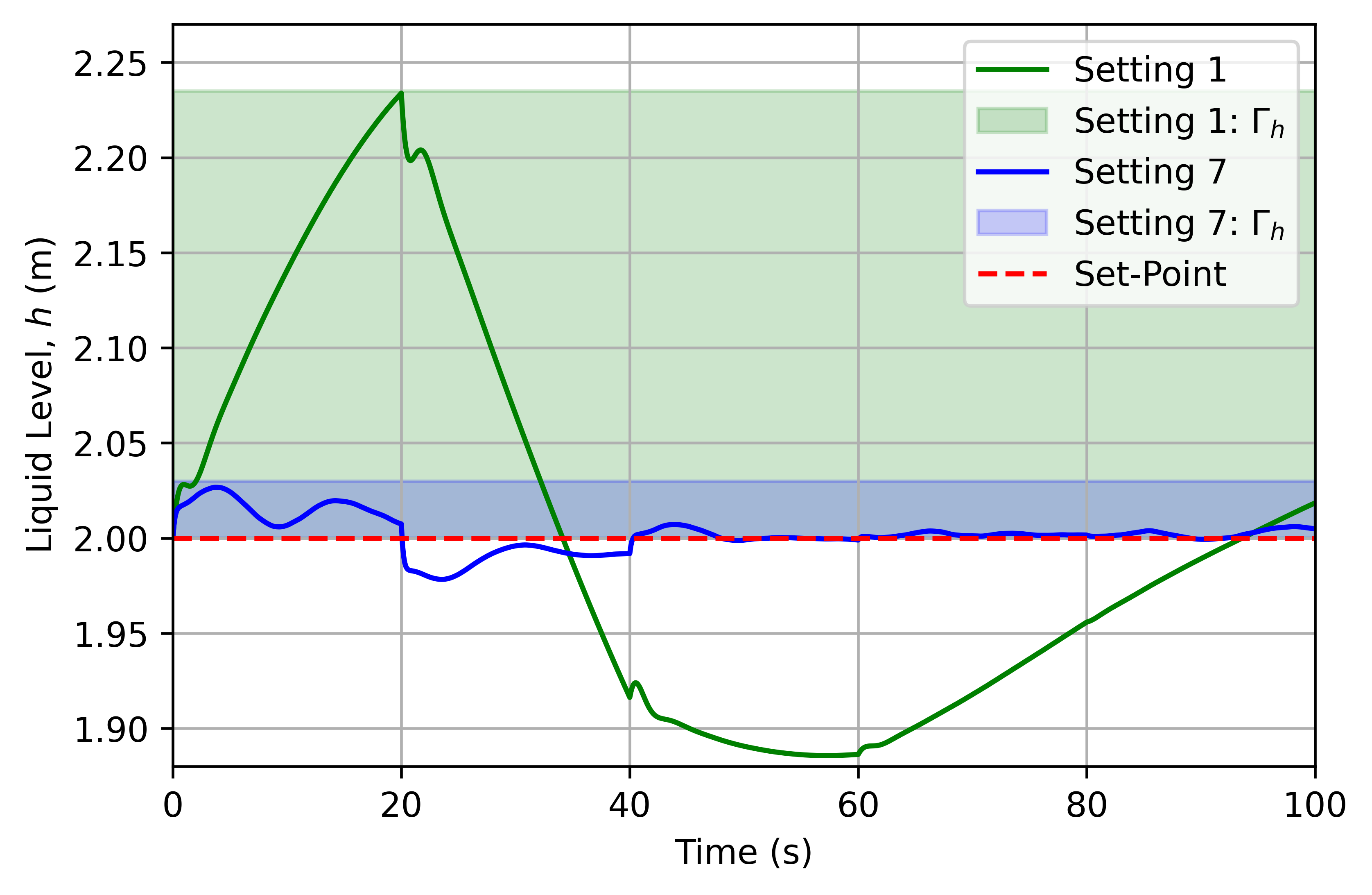}
         \caption{}
         \label{fig:level_control}
     \end{subfigure}
     \begin{subfigure}[b]{0.32\textwidth}
         \centering
         \includegraphics[width=\textwidth]{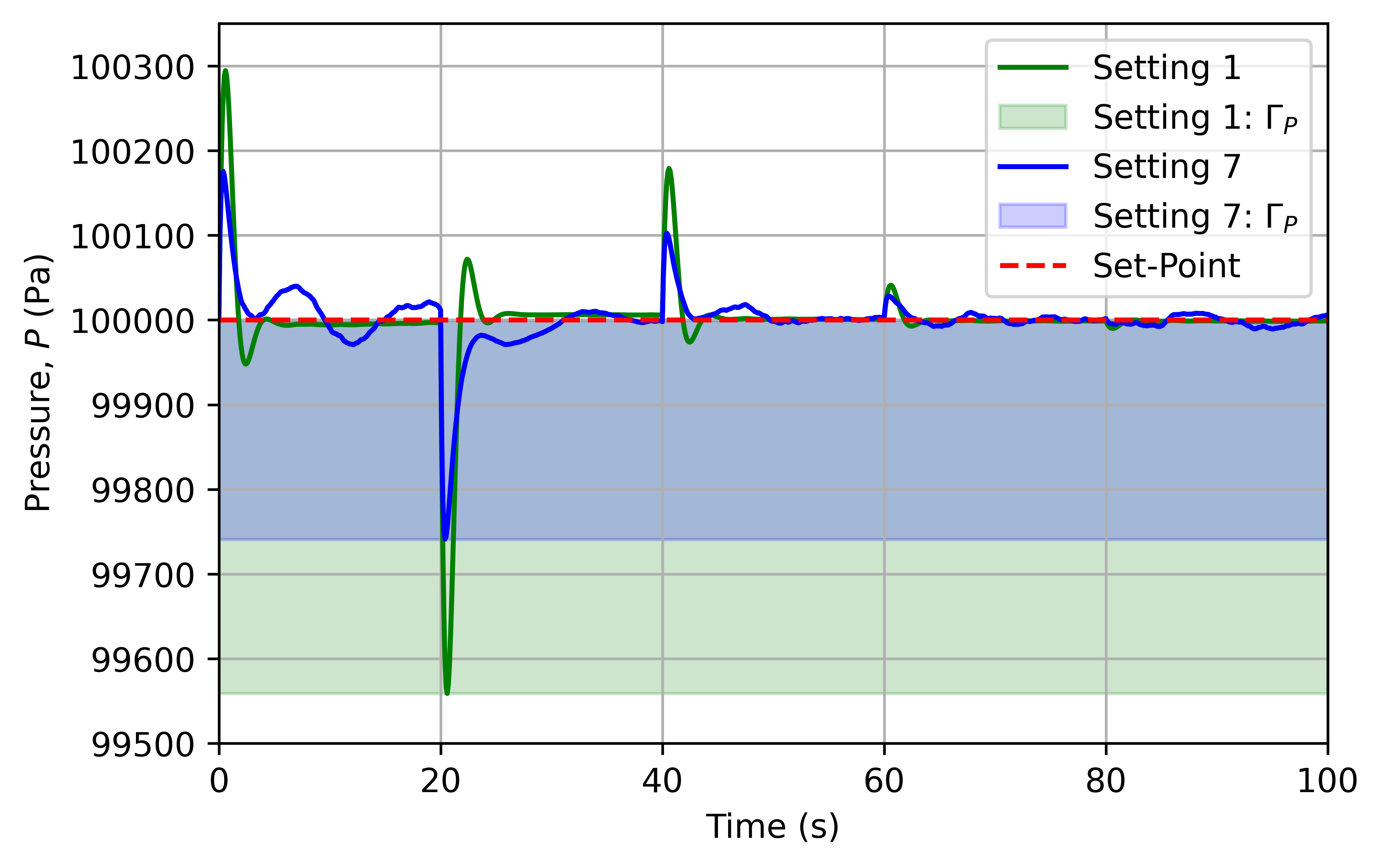}
         \caption{}
         \label{fig:pressure_control}
     \end{subfigure}
        \caption{Controller tuning results for setting 1 (green) and setting 7 (blue), with set-points indicated by red dashed horizontal lines. Coloured hues represent the maximum state perturbations from the set-point, used as $\Gamma_i$ values in the ARRTOC algorithm. (a): Product composition control. (b): Liquid level control. (c): Pressure control.}
        \label{fig:controller_tuning}
\end{figure}

After this, the best gains found for each of the controller settings were then used to determine the largest perturbation the controllers could handle for each of the states. This was again tested around the arbitrarily chosen set-point. These values were then used as a proxy measure to define the level of robustness desired at the RTO layer via the ARRTOC parameters: $\Gamma_{P}, \Gamma_{h}$ and $\Gamma_{x_B}$. A margin of error could be incorporated here to accommodate the tuning uncertainties discussed earlier. Notably, this step benefits from the general uncertainty set defined in equation (\ref{eq:ellipsis_uncertainty_set}), due to the asymmetric effect of the disturbances on each of the states as a result of the control loop interactions. Table \ref{param_table_4} presents the seven controller designs, their associated gains, summed normalised ITAE value, and the largest perturbation each of the states experience from the set-point. As an example, we also show in Figure \ref{fig:controller_tuning} the controller performance of controller settings 1 and 7 with the maximum state perturbations (i.e. the values of $\Gamma_{P}, \Gamma_{h}$ and $\Gamma_{x_B}$) depicted as a green (setting 1) and blue (setting 7) hue respectively. 

Next, for each of the controller settings, the constrained ARO equivalent of the nominal problem in equation (\ref{eq:evaporator_nominal}) was solved using the ARRTOC algorithm from section \ref{sec:background_meth} with the uncertainty set given by:

\begin{equation}\label{evaporator_uncertainty_set}
\mathcal{U} = \left\{\boldsymbol{\Delta}\mathbf{x} \; \middle | 
 \; \frac{{\Delta P}^2}{{\Gamma_P}^2} + \frac{{\Delta h}^2}{{\Gamma_h}^2} + \frac{{\Delta x_B}^2}{{\Gamma_{x_B}}^2} \leq 1 \right\}
\end{equation}
where, $\boldsymbol{\Delta}\mathbf{x} = [\Delta x_B, \Delta h, \Delta P]^T \in \mathbb{R}^3$, are the possible implementation errors of the system. The values of $\Gamma_i$ for each of the states and controller settings is shown in Table \ref{param_table_4}. Moreover, the adversarially robust optimum set-points and profit are shown in the final 2 columns of the table. 

\begin{table}
\caption{Table of controller tuning and ARRTOC results.}
\normalsize
\begin{center}
\resizebox{\textwidth}{!}{\begin{tabular}{c|p{0.08\textwidth}p{0.08\textwidth}p{0.08\textwidth}p{0.08\textwidth}p{0.08\textwidth}p{0.08\textwidth}p{0.08\textwidth}p{0.08\textwidth}p{0.05\textwidth}p{0.05\textwidth}p{0.05\textwidth}p{0.19\textwidth}p{0.05\textwidth}}
\hline
& BOBYQA Budget, $N$& $KC_{PD}$ & $KI_{PD}$ & $KC_{hB}$ ($\times10^3$) & $KI_{hB}$ & $KC_{x_BT_S}$ ($\times10^3$)  & $KI_{x_BT_S}$ & ITAE & $\Gamma_P$ (Pa) & $\Gamma_h$ (m) & $\Gamma_{x_B}$ (-) & ARO S.P\newline($x_B^*$, $h^*$, $P^*$) & ARO Profit (\$/sec) \\ \hline
\textbf{Setting 1} & $10$ & $0.10$ & $0.20$ & $0.10$ & $2.50$ & $0.10$ & $0.50$ & $3142.17$ & $441$ & $0.23$ & $0.13$ & (0.77, 2.23, 0.100441) & $58.08$ \\
\textbf{Setting 2} & $25$ & $0.05$ & $0.40$ & $0.50$ & $1.25$ & $0.05$ & $0.25$ & $2700.24$ & $394$ & $0.25$ & $0.16$ & (0.74, 2.25, 0.100394) & $51.08$ \\
\textbf{Setting 3} & $50$ & $0.10$ & $0.20$ & $0.10$ & $2.50$ & $1.00$ & $0.50$ & $2488.29$ & $457$ & $0.17$ & $0.08$ & (0.82, 2.17, 0.100457) & $69.85$ \\
\textbf{Setting 4} & $100$ & $0.20$ & $0.10$ & $0.05$ & $1.25$ & $0.50$ & $1.00$ & $2350.84$ & $339$ & $0.20$ & $0.05$ & (0.85, 2.20, 0.100339) & $76.73$ \\
\textbf{Setting 5} & $500$ & $0.19$ & $0.11$ & $0.07$ & $1.10$ & $1.37$ & $0.00$ & $2032.22$ & $322$ & $0.11$ & $0.06$ & (0.84, 2.11, 0.100322) & $74.70$ \\
\textbf{Setting 6} & $750$ & $0.10$ & $0.10$ & $2.50$ & $5.07$ & $1.00$ & $0.50$ & $1817.36$ & $309$ & $0.02$ & $0.04$ & (0.86, 2.02, 0.100309) & $79.63$ \\
\textbf{Setting 7} & $1000$ & $0.19$ & $0.17$ & $1.25$ & $6.37$ & $1.28$ & $0.47$ & $1783.32$ & $259$ & $0.03$ & $0.04$ & (0.86, 2.03, 0.100259) & $79.61$ \\ \hline
\end{tabular}}
\label{param_table_4}
\end{center}
\end{table}

\begin{figure}
     \centering
     \begin{subfigure}[b]{0.45\textwidth}
         \centering
         \includegraphics[width=\textwidth]{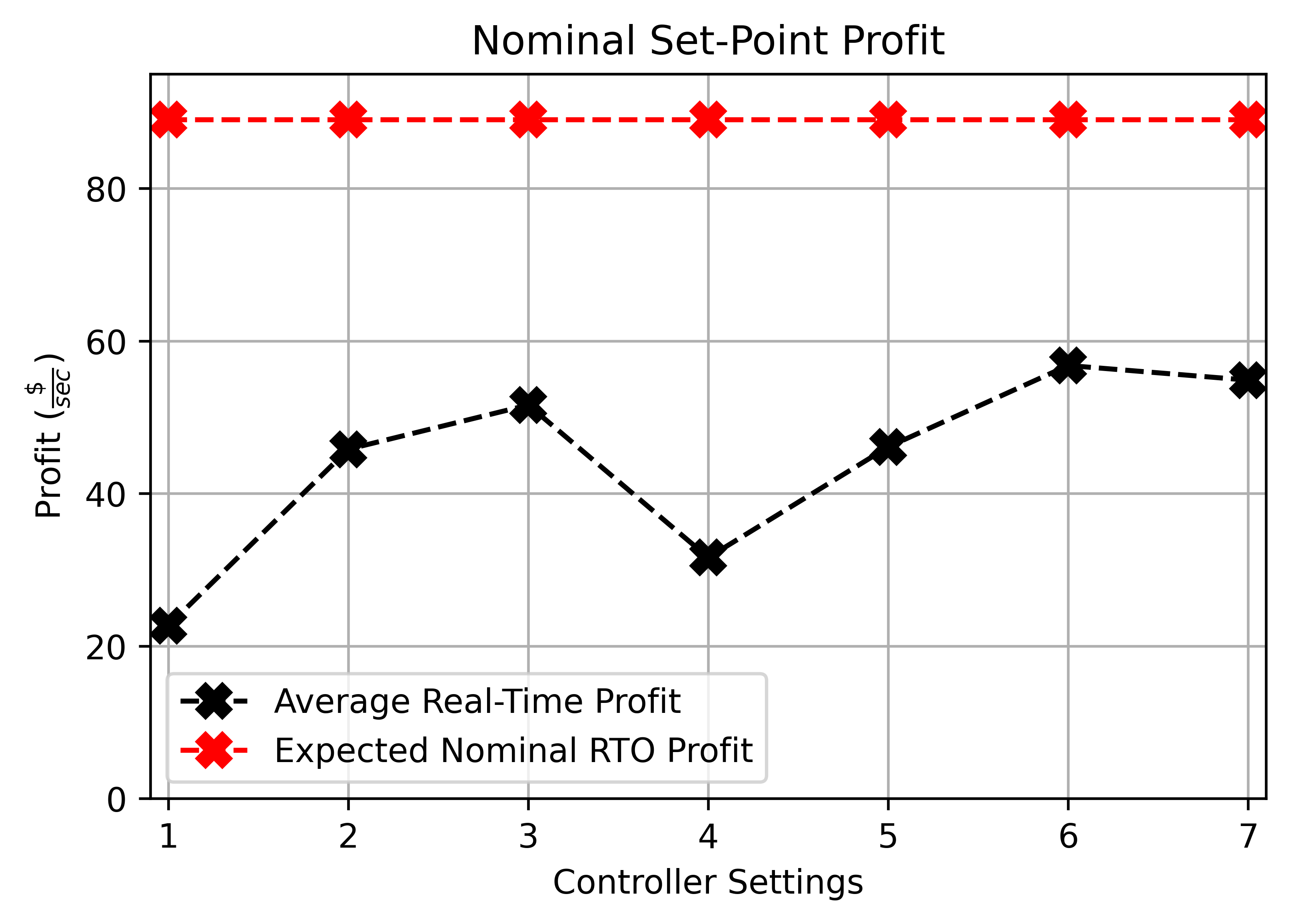}
         \caption{}
         \label{fig:controller_nominal_profit}
     \end{subfigure}
     \begin{subfigure}[b]{0.45\textwidth}
         \centering
         \includegraphics[width=\textwidth]{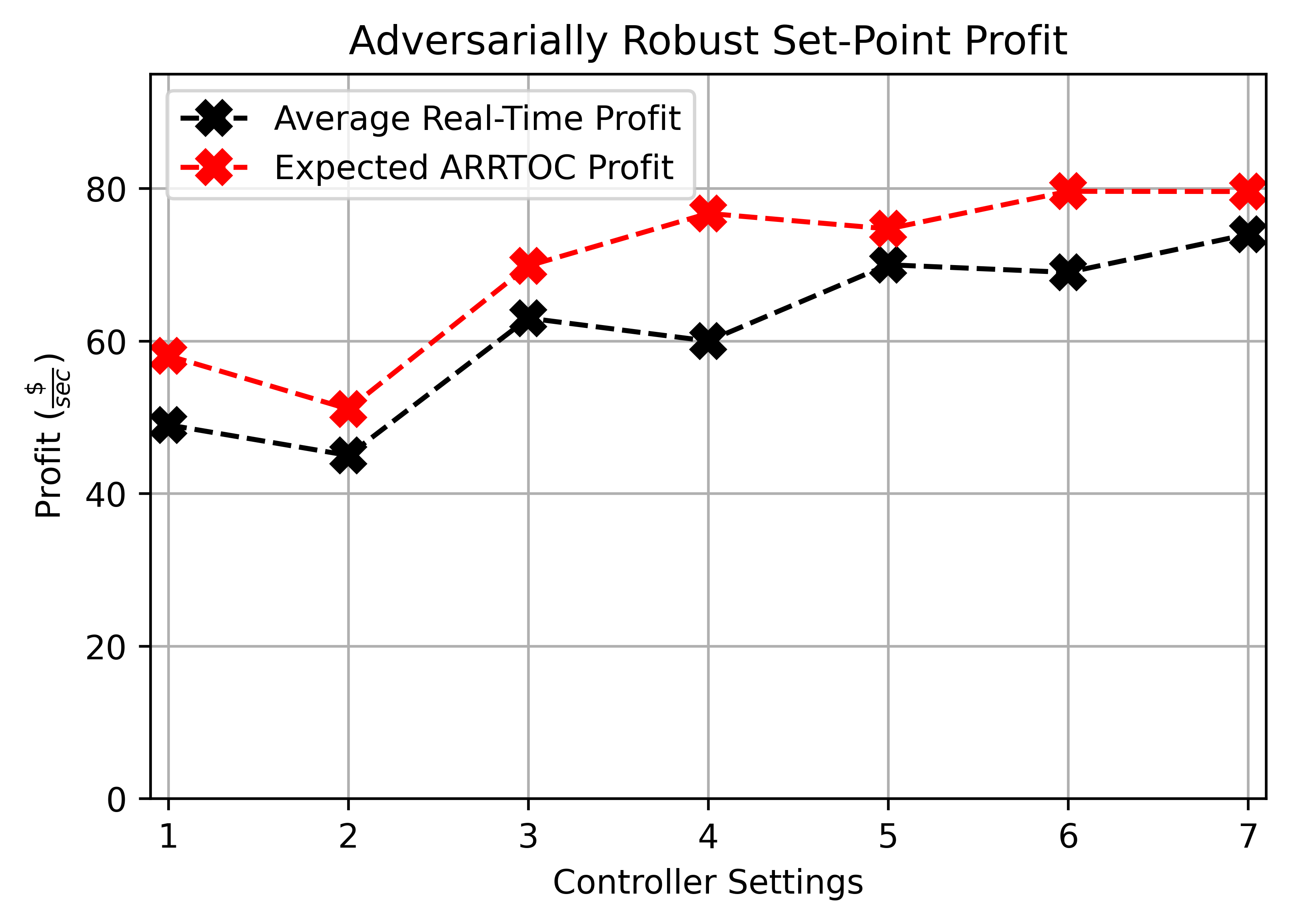}
         \caption{}
         \label{fig:controller_robust_profit}
     \end{subfigure}
        \caption{(a): Average real-time profit (black crosses) compared to expected profit at the nominal optimum (red crosses) for each controller setting. (b): Expected adversarially robust optimum profit (red crosses) contrasted with average real-time profit (black crosses) for each setting operating around their respective adversarially robust set-points.}
        \label{fig:profit_comparison}
\end{figure}

\begin{figure}
     \centering
     \begin{subfigure}[b]{0.45\textwidth}
         \centering
         \includegraphics[width=\textwidth]{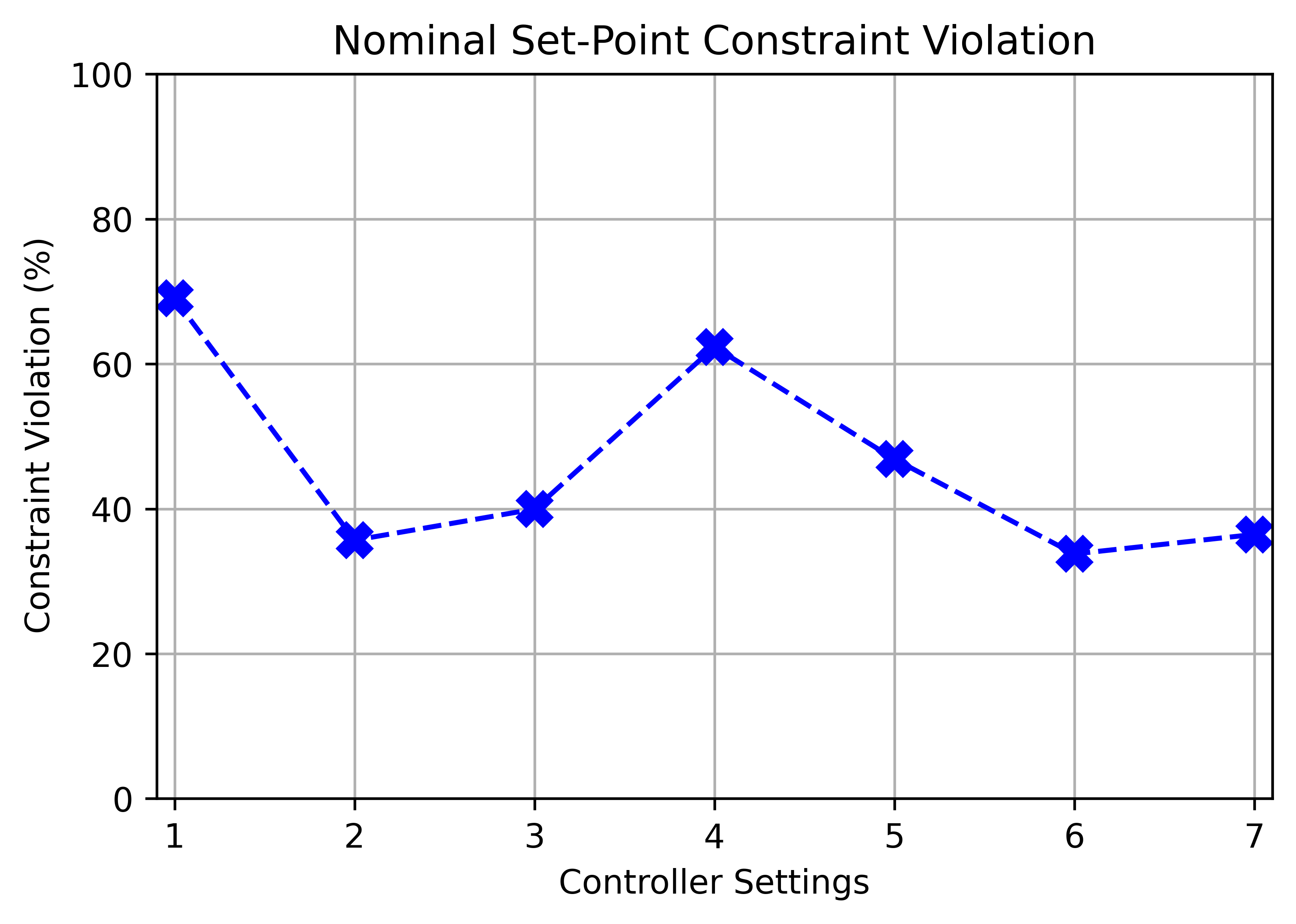}
         \caption{}
         \label{fig:controller_nominal_violation}
     \end{subfigure}
     \begin{subfigure}[b]{0.45\textwidth}
         \centering
         \includegraphics[width=\textwidth]{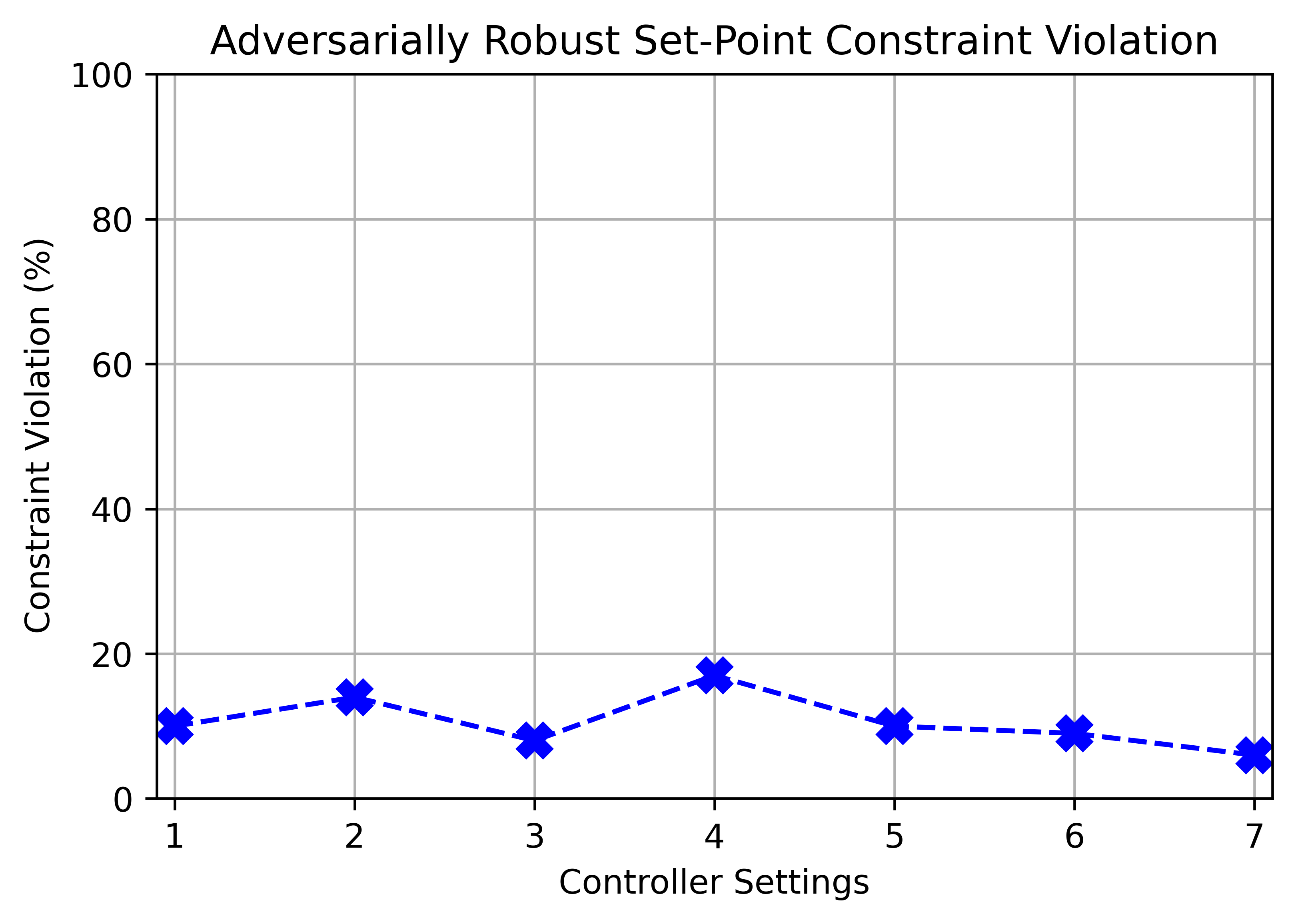}
         \caption{}
         \label{fig:controller_robust_violation}
     \end{subfigure}
        \caption{(a): Constraint violation percentage for each setting operating around the nominal optimum set-point. (b): Constraint violation percentage for each setting operating around their respective adversarially robust set-points.}
        \label{fig:violation_comparison}
\end{figure}

Finally, given these 7 different controller designs we test the performance of the controllers when operating at their respective adversarially robust optima as compared to the benchmark nominal optimum. We assume the same disturbance characteristics as before. For both scenarios, we record the real-time profit during the length of the simulation and if any constraints are violated. For any timesteps where constraints are violated, we assume the instantaneous profit value is $0$. We then average the real-time profit for the length of the simulation and evaluate the percentage of time the system is in a constraint violating state. The results of this are depicted in Figures \ref{fig:profit_comparison} and \ref{fig:violation_comparison}. 

Figure \ref{fig:profit_comparison} shows the average profit for each of the controller settings when operating around the nominal optimum set-point (Figure \ref{fig:controller_nominal_profit}) and the respective adversarially robust optimum set-points (Figure \ref{fig:controller_robust_profit}). In both figures, the red crosses depict the expected optimum profit at the set-point with Figure \ref{fig:controller_nominal_profit} showing the expected nominal profit and Figure \ref{fig:controller_robust_profit} showing the expected ARRTOC profit. The black crosses show the actual average real-time profit delivered for each setting i.e. under the action of control. The gap in optimality in both cases arises from the fact that the three controllers attempt to keep the controlled states at their respective set-points but due to system disturbances (Figure \ref{fig:disturbance_signals}) exact control at the set-point can not be achieved. Consequently, the average real-time profit is lower than the expected optimal profit. However, clearly there is a stark difference between operating at the nominal set-point and the respective adversarially robust set-points. Specifically, it is clear to see in Figure \ref{fig:controller_nominal_profit}, there is a significant optimality gap between the expected nominal profit and the actual average real-time profit. Generally, the controller performance, as measured by the average real-time profit, only improves when the controller robustness improves. Effectively, the nominal set-point completely ignores the underlying controller designs and implicitly assumes all of the controllers are highly robust. This means that if the system or controllers were to degrade over time, which is likely to occur, then the nominal set-point would be a wholly unacceptable choice. On the other hand, in Figure \ref{fig:controller_robust_profit}, the gap in optimality is significantly smaller in comparison. Generally, by considering the underlying controller design, the ARRTOC algorithm finds set-points which account for the robustness already available at the control layer. For instance, for controller setting 1 which has the worst robustness to disturbances compared to the other designs, the expected adversarially robust profit of $\frac{\text{\$}}{\text{s}}58.08$ is approximately $35\%$ lower than the expected nominal profit of $\frac{\text{\$}}{\text{s}}89.03$. At face value, this appears to be a major disadvantage until we compare the actual profit delivered. For controller setting 1 from Figure \ref{fig:controller_nominal_profit}, the average real-time profit is $\frac{\text{\$}}{\text{s}}23.01$ compared to the same setting from Figure \ref{fig:controller_robust_profit} where the average real-time profit is $115\%$ larger at $\frac{\text{\$}}{\text{s}}49.55$.

Figure \ref{fig:violation_comparison} shows the constraint violation percentage for each controller setting around the nominal optimum set-point (Figure \ref{fig:controller_nominal_violation}) and the corresponding adversarially robust optimum set-points (Figure \ref{fig:controller_robust_violation}). The observations mirror those of Figure \ref{fig:profit_comparison}. In Figure \ref{fig:controller_nominal_violation}, the nominal optimum set-point proves to be an extremely poor choice from an operability perspective. Even the best controller setting experiences constraint violations for approximately $40\%$ of the simulation time, while the least robust controller setting violates constraints for nearly $70\%$ of the time. These results underscore the poor performance of the nominal set-point, which fails to account for the varying levels of robustness among the controller settings. Conversely, in Figure \ref{fig:controller_robust_violation}, the constraint violation percentages exhibit minimal variation, with an average hovering around $10\%$. Notably, the highest constraint violation percentage recorded is slightly above $15\%$ for controller setting 4, which significantly outperforms the best-case scenario presented in Figure \ref{fig:controller_nominal_violation}. These findings highlight the effectiveness of the ARRTOC algorithm in optimizing control system operability. By accounting for the robustness of individual controller settings through tailored set-points, ARRTOC ensures that constraints are consistently adhered to, with minimal violations across the board. As a final point, it is important to note that the ARRTOC robustness parameters, detailed in Table \ref{param_table_4}, were derived specifically from controller tunings around the chosen set point of $(x_B = 0.7, h = 5\text{m}, P = 0.1\text{MPa})$. When applied to different set-points, as in this case, these robustness parameters may not be the "best". However, as previously discussed, this situation allows for the inclusion of an additional safety margin to the robustness parameters or they could be re-estimated around the newly chosen set-points. This explains why constraint violations do not reach $0\%$ in Figure \ref{fig:controller_robust_violation}. Notably, our findings illustrate that despite potentially sub-optimal choices for the robustness parameters, excellent performance can still be attained compared to the nominal case.

To demonstrate the feasibility of incorporating ARO at the RTO layer, we conclude this section by providing a brief computational comparison between the nominal RTO problem and the adversarially robust RTO problem for this case study. All computations were performed on a machine with 2 AMD EPYC 7742 processors. The nominal RTO problem was solved using the IPOPT algorithm and a solution was found in 1.47 seconds on average. In contrast, the ARRTOC algorithm took roughly 5 times longer on average with a solution being found in approximately 7.10 seconds across the different controller settings considered. Profiling the ARRTOC algorithm revealed that approximately 82\% of the computational time was allocated to the neighbourhood exploration routines (sections \ref{constrained:step_1} and \ref{constrained:step_2}). This is due to the fact that the neighbourhood exploration routines are multi-start and rely on first-order information alone, which could be improved by second-order information availability to enhance efficiency. However, our goal was to provide an algorithm which could be easily adopted for various RTO problem settings including those reliant on function evaluations and gradients alone, typical of process simulation models. In scenarios where these restrictions do not apply, and second-order information is readily available then the neighbourhood exploration steps can be adapted to exploit this information, thus reducing the computational demand. On the other hand, the remainder of the ARRTOC algorithm is allocated to the robust local move routines (sections \ref{constrained:step_3a} and \ref{constrained:step_3b}). These routines are extremely efficient as they are primarily concerned with solving a convex second order cone program which has limited computational cost in comparison to the multi-start gradient ascent neighbourhood explorations. Fundamentally, as expected, while the robust problem incurs greater computational overhead, the RTO layer is better able to absorb this cost in comparison to the control layer where such increases in computational demand would not be acceptable. By using the greater computational time resources available at the RTO layer to find robust set-points for the control layer, we can reduce the demand on the resource-limited controllers while also improving performance as we have demonstrated in this case study.

\section{Conclusions and Future Work}\label{sec:conclusion}

The ARRTOC algorithm has showcased its ability to address the dual challenges of operability and optimality, focussing on the robust handling of implementation errors such as disturbances and noise at the RTO layer. By leveraging the robustness embedded in the controller design, ARRTOC has emerged as a solution for finding tailored set-points that are resilient to implementation errors at the control layers.

In section \ref{sec:background_meth}, we provided a tutorial-style introduction to the ARRTOC algorithm. Thereafter, the illustrative example presented in section \ref{rd:illustrative_example} shed light on several key insights of the ARRTOC algorithm. Notably, it highlighted the benefits of adversarially robust set-points in improving operability at the control layers. Additionally, the importance of integrating controller design robustness into ARRTOC was demonstrated, emphasising the need for set-points that strike a balance between robustness at both layers. Thereafter, we leveraged these insights and applied them to the practical case studies of section \ref{rd:bioreactor} and \ref{rd:evaporator}. In section \ref{rd:bioreactor}, the algorithm's ability to address the challenge of continuous bioprocess operation and prevent washout was demonstrated. Additionally, ARRTOC was shown to be compatible with various underlying model structures, whether data-driven or mechanistic. In section \ref{rd:evaporator}, the insights from the illustrative example were successfully applied to a multi-loop evaporator process where we used the controller design to inform the level of robustness desired at the RTO layer. This approach resulted in tailored set-points that were neither excessively conservative nor overly risky. Simulation results confirmed the consistent and favourable performance of the system, regardless of the underlying controller design. 

Finally, while ARRTOC has already shown its effectiveness in addressing implementation errors, there are exciting directions for future development. By integrating adaptation algorithms to handle plant-model mismatch, a major challenge in RTO \cite{Ahmed2021, Zagorowska2023, Chanona2021, Chachuat2009, Mercangoz2008}, and exploring different controller structures such as MPC, ARRTOC can evolve into a comprehensive and practical solution for robust real-time optimization and control. Empirical validation through lab-scale and pilot-scale experiments will further solidify its advantages and position ARRTOC as a valuable and pragmatic tool for such problems. 

\bibliographystyle{ieeetr}
\begin{spacing}{1}
\bibliography{arrtoc}  
\end{spacing}

\end{document}